%% file: freq-banded_SGWB.tex
\documentclass[%
reprint,
preprintnumbers,
showkeys,
nofootinbib,
 amsfonts, amsmath,amssymb,
aps,
prd,
longbibliography
]{revtex4-2}

\usepackage{amsthm}
\usepackage{mathtools}
\usepackage{physics}
\usepackage{xcolor, soul, ulem}
\usepackage{adjustbox}
\usepackage{placeins}
\usepackage[T1]{fontenc}
\usepackage{blindtext}
\usepackage[caption=false]{subfig}

\usepackage[english]{babel}
\usepackage[utf8]{inputenc}
\usepackage{graphicx}
\usepackage{dcolumn}
\usepackage{bm}
\usepackage[pdftex, pdftitle={Article}, pdfauthor={Author}]{hyperref}
\usepackage[mathlines]{lineno}

\graphicspath{{figures/}} 

\newcommand*\diff{\mathop{}\!\mathrm{d}}

\begin{document}


\title{Model-independent search for anisotropies in stochastic gravitational-wave backgrounds and application to LIGO--Virgo's first three observing Runs}

\author{Liting Xiao}
    \email{lxxiao@caltech.edu}
    \affiliation{LIGO Laboratory, California Institute of Technology, Pasadena, CA 91125, USA}

\author{Arianna I. Renzini}
    \email{arenzini@caltech.edu}
    \affiliation{LIGO Laboratory, California Institute of Technology, Pasadena, CA 91125, USA}

\author{Alan J. Weinstein}
    \affiliation{LIGO Laboratory, California Institute of Technology, Pasadena, CA 91125, USA}

\begin{abstract}
    A stochastic gravitational-wave (GW) background consists of a large number of weak, independent and uncorrelated events of astrophysical or cosmological origin. The GW power on the sky is assumed to contain anisotropies on top of an isotropic component, i.e., the angular monopole. Complementary to the LIGO--Virgo--KAGRA (LVK) searches, we develop an efficient analysis pipeline to compute the maximum-likelihood anisotropic sky maps in stochastic backgrounds directly in the sky pixel domain using data folded over one sidereal day. We invert the full pixel-pixel correlation matrix in map-making of the GW sky, up to an optimal eigenmode cutoff decided systematically using simulations. In addition to modeled mapping, we implement a model-independent method to probe spectral shapes of stochastic backgrounds. Using data from LIGO--Virgo's first three observing runs, we obtain upper limits on anisotropies as well as the isotropic monopole as a limiting case, consistent with the LVK results. We also set constraints on the spectral shape of the stochastic background using this novel model-independent method.

\end{abstract}


\maketitle

\input{sections/1_intro}
\input{sections/2_modeled_method}
\input{sections/3_unmodeled_method}
\input{sections/4_simulations}
\input{sections/5_LIGO-Virgo_data}
\input{sections/6_conclusions}
\input{sections/7_acknowledgements}

\bibliography{references}

\end{document}

%% file: sections/1_intro.tex
\section{Introduction} \label{sec:intro}

Direct detections of gravitational waves (GWs) from Advanced LIGO \cite{aLIGO_det}, Advanced Virgo \cite{Virgo_det} and KAGRA \cite{KAGRA_det} detectors so far have been traced back to point-like sources, which make up a tiny fraction of the GW sky. The bulk of unresolved signals associated with multiple point sources or extended sources combine incoherently to form backgrounds of GWs. A stochastic gravitational-wave background (SGWB) consists of a large number of independent and uncorrelated events which are typically individually weak, i.e., below the detection threshold of the detector. SGWBs can be categorized as either astrophysical, when produced by low-redshift, individually indistinguishable GW events \cite{Regimbau_astro_sources, Regimbau_astro_SGWB}, or cosmological \cite{cosmo_SGWB_review}, as a result of high-energy events in the early Universe such as cosmic inflation \cite{inflation_SGWB1, inflation_SGWB2, inflation_SGWB3}, cosmic string networks \cite{cosmic_string_lecture, GW_cusp_kink, GW_cusp_kink2, SGWB_cosmic_string}, primordial black holes \cite{SGWB_primordialBH, SGWB_primordialBH2, SGWB_primordialBH3}, and first order phase transitions \cite{SGWB_phase_trans, SGWB_phase_trans2, SGWB_bubble, SGWB_bubble2, SGWB_magnet, SGWB_sound}.

Ground-based GW detectors are sensitive to SGWBs constrained between tens of Hz and a few hundred Hz. In other frequency bands, upper limits on SGWBs are set by the isotropy of the Cosmological Microwave Background (CMB) \cite{CMB_Planck2018_isotropy} in the lowest frequencies \cite{CMB_wmap_bicep_keck}, by timing residual analyses in millisecond pulsars in the nHz band \cite{nanograv_SGWB}, by normal modes of the Earth \cite{Earth_normal_modes} and the Moon \cite{Moon_normal_modes} in the mHz to Hz band, and loosely by primordial deuterium abundance from Big Bang Nucleosynthesis \cite{early_universe, GWB_BBN} over a broad frequency range.

Studying SGWBs may potentially open a window onto ~$\sim10^{-32}$ s (at a redshift $z > 10^{25}$) after the Big Bang. Our current knowledge of the early Universe mostly comes from the CMB \cite{CMB_og, CMB_review}, the relic electromagnetic (EM) radiation from 380,000 years ($z \sim 1100$) after the Big Bang during the epoch of recombination. Before recombination, the Universe was opaque to EM waves. Hence, GWs present a unique opportunity to probe the earliest moments of the Universe.

Background gravitational radiation is stochastic in the sense that it can only be characterized statistically, in terms of moments of its probability distribution. Stochastic GW signals can mimic shot noise, appearing as individual bursts in the timestream, or they can be described as Gaussian, where a multitude of signals overlap so that the central limit theorem applies. They may also exhibit popcorn-like features in the time domain, with partially overlapping signals but still far from the Gaussian regime~\cite{SGWB_review2022_AR}. To differentiate between the aforementioned sources of stochastic backgrounds, several subtraction or multi-fitting methods have been proposed~\cite{Biscoveanu:2020gds,Zhong:2022ylh,Zhou:2022otw}, which leverage on the particular statistical nature of each signal contribution.

At cosmological scales, we assume the GW sky is isotropic based on the isotropy of the CMB, which traces the matter distribution in the Universe. However, at local scales, the nonuniform distribution of astrophysical GW sources may produce an anisotropic SGWB. Moreover, similar to the CMB dipole anisotropy \cite{CMB_anisotropies_review, CMB_Planck2018_anisotropies, CMB_dipole_contestation}, 
our peculiar motion with respect to the SGWB rest frame induces a recurring modulation affecting the dipole. It is thus fair to assume the SGWB power contains anisotropic components on top of an ensemble average isotropic value.

An approach to reconstruct the angular power distribution in an anisotropic SGWB is computing the maximum-likelihood map solutions using cross-correlated data \cite{cross_corr_Christensen, cross_corr_Flanagan, cross_corr_Michelson} from a network of ground-based GW detectors. This is typically done assuming stationary Gaussian detector noise and a specific model for the spectral power distribution of the signal, and employing a weak-signal approximation~\cite{LIGO_radiometer_method2, SHD_method}. The latter implies we can ignore any signal contribution to the data auto-correlations, essentially allowing us to estimate the noise covariance from the data directly~\cite{Matas_Romano_21}. Mapping can be carried out in any set of basis functions on the sky, e.g., spherical harmonics of the SGWB power as adopted by the LIGO Scientific, Virgo and KAGRA (LVK) collaboration \cite{O3_SGWB_isotropic, O3_SGWB_anisotropic}, or sky pixel indices as in \cite{LIGO_first_pixel, AR_method, AR_O1, AR_O1O2, Sambit_mapping, Deepali_fisher}.

Directional searches by the LVK collaboration \cite{PhysRevLett.107.271102, O1_anisotropic, O2_anisotropic, O3_SGWB_anisotropic} include the broadband radiometer analysis (BBR) \cite{LIGO_radiometer_method}, the spherical harmonic decomposition (SHD) \cite{SHD_method}, the narrow band radiometer analysis (NBR) \cite{LIGO_radiometer_method}, and the all-sky, all-frequency analysis (ASAF) \cite{O3_ASAF}. The BBR targets persistent point sources emitting GWs over a wide frequency range, whereas the SHD hunts for extended sources with smooth frequency spectra. The NBR inspects frequency spectra from specific locations on the sky, such as Scorpius X-1, SN 1987A and the Galactic Center, in narrow frequency bands. The ASAF scans the sky in individual frequency bins, searching for excess GW power for each pixel-frequency pair. The BBR, SHD and NBR approaches integrate over frequencies employing a filter which includes a power-law model for the SGWB power spectrum, while the ASAF is a generic unmodeled search. Out of all of these, the SHD search is the only one that takes pixel-pixel correlations into account.


Complementary to the LVK searches, we present an efficient analysis pipeline built in \texttt{Python} to map anisotropies in SGWBs directly in the sky pixel domain using data folded over one sidereal day. Our pipeline is tailored to folded data  \cite{data_folding_fast, data_folding_veryfast, Boris_folding}, which assumes the SGWB signal is stationary (i.e., is time-independent) and exploits the temporal symmetry of the Earth's rotation to reduce the computation time by a factor of total observing days. In the pipeline, we use the \texttt{HEALPix} hierarchical pixelization scheme \cite{healpix}, in which the sky is discretized into equal area elements. We invert the ``full'' pixel-pixel correlation matrix in map-making of the GW sky, up to an optimal eigenmode cutoff decided systematically using simulations. In addition to modeled maximum-likelihood mapping, we implement a spectral-model-independent method to probe the spectral shape of a SGWB based on previous work in \cite{AR_O1O2}, now taking into account the deconvolution regularization problem systematically as a function of frequency. In both approaches, sky maps are converted from the pixel domain to the Fourier domain to place upper limits on the angular power spectrum, as well as the power spectrum of the monopole component of the background.

In Sec.~\ref{sec:method}, we outline our methodology to compute the maximum-likelihood map solutions of the SGWB sky power assuming a standard model for the power spectrum. In Sec.~\ref{sec:gwb_model_indep} we outline our unmodeled approach to mapping, where we employ adaptive frequency-banding and an adaptive pixelization scheme to probe the shape of the signal power spectrum, as well as recovering the angular distribution. In Sec.~\ref{sec:simulations}, we describe the various simulations used to verify our pipeline. In Sec.~\ref{sec:data}, we apply our pipeline to data from LIGO--Virgo's first three observing runs (O1-O3). In Sec.~\ref{sec:conclusions}, we discuss our results and outlook for upcoming observing runs and the field of GW cosmology.

%% file: sections/2_modeled_method.tex
\section{SGWB Mapping} \label{sec:method}

\subsection{Energy Density Spectrum}

A SGWB is characterized by its spectral emission. Specifically, we introduce a dimensionless quantity, the normalized GW energy density spectrum,
\begin{equation} \label{eq:SGWB_spectrum}
    \Omega_\mathrm{GW}(f) \equiv \frac{1}{\rho_c} \frac{\diff \rho_\mathrm{GW}}{\diff \log f},
\end{equation}
where $\rho_\mathrm{GW}$ is the GW energy density and $\rho_c$ is the critical energy density required to close the Universe today, 
\begin{equation}
    \rho_c = \frac{3 H_0^2 c^2}{8 \pi G}.
\end{equation}
Here, $c=2.998\times10^8$ m s$^{-1}$ is the speed of light and $H_0 = 67.4$ km s$^{-1}$ Mpc$^{-1}$ \cite{CMB_Planck2018_Hubble} is the Hubble expansion rate of the current epoch (with some controversy on its measured value in the literature \cite{Hubble_constant_review, Hubble_TRGB, CMB_Planck2018_Hubble, gwtc3_hubble}). 
Conceptually, $\Omega_\mathrm{GW}(f)(\diff f/f)$ is the ratio of the GW energy density to the total energy density required to close the Universe today in a small frequency interval from $f$ to $f + \diff f$.

$\Omega_{\rm GW}(f)$ is a sky-averaged quantity, and may be written as an integral over the sky of the directional energy density $\Omega_{\rm GW}(f, \Theta)$,
\begin{equation}\label{eq:anisotropic_omega_integral}
   \Omega_\mathrm{GW}(f) = \frac{1}{4 \pi} \int_{S^2} \diff \Theta \, \Omega_\mathrm{GW}(f, \Theta),
\end{equation}
where $\Theta$ is a direction on the sky on the two sphere in a general basis. $\Omega_\mathrm{GW}(f, \Theta)$ may be interpreted as the energy density spectrum in each direction, and is the target of several anisotropic stochastic background searches. As our detectors measure GW strain, it is useful to report the relation between the energy density and the GW strain power $\mathcal{P}(f, \Theta)$,
\begin{equation}\label{eq:anisotropic_omega_strain}
   \Omega_\mathrm{GW}(f, \Theta) = \frac{4\pi^2}{\rho_c G} f^3 \mathcal{P}(f, \Theta).
\end{equation}
This follows directly from Isaacson's formula for GW radiation, which implies~\cite{PhysRev.166.1263, SGWB_review2022_AR}
\begin{equation}
    \rho_{\rm GW} = \frac{\pi}{G} \int_0^{\infty} \diff f \int \diff\Theta \, f^2 {\cal P}(f, \Theta) .
\end{equation}
Different conventions are used at times when defining the normalization of the quantities above. We employ the conventions as in~\cite{SGWB_review2022_AR}.

For the sake of simplicity, stochastic searches typically assume that the directionality and the spectral shape of the signal are independent, such that the GW strain power $\mathcal{P}(f, \Theta)$ in Eq.~\eqref{eq:anisotropic_omega_strain} may be factored into a spectral term and an angular term\footnote{See Sec.~\ref{sec:gwb_model_indep} for a brief discussion of the validity of this assumption.},
\begin{equation}\label{eq:anisotropic_power_law}
    \mathcal{P}(f, \Theta) = H(f)\mathcal{P}(\Theta).
\end{equation}

The spectral shape $H(f)$ is usually modeled as a power law given by
\begin{equation}\label{eq:spectral_shape}
    H_\alpha(f) = \bigg( \frac{f}{f_\mathrm{ref}} \bigg) ^ {\alpha-3},
\end{equation}
where $\alpha$ is the spectral index and $f_\mathrm{ref}$ is a reference frequency. %
This choice of model is well-motivated by many astrophysical and cosmological models~\cite{Regimbau_astro_SGWB, cosmo_SGWB_review}, however there are well-known spectral shapes outside this regime, e.g., the combined SGWB from compact binary coalescences (CBCs) at higher frequencies~\cite{gwtc3_pop}. The power-law assumption is a good approximation for the CBC SGWB at current detector sensitivities, but is expected to break down as sensitivity increases. %
For a direct comparison with the LVK results \cite{O3_SGWB_anisotropic}, we also set $f_\mathrm{ref}$ to 25 Hz.

$\mathcal{P}(\Theta)$ in Eq.~\eqref{eq:anisotropic_power_law} is the angular power distribution that can be expanded in a set of chosen basis functions $e_\eta(\Theta)$ on the two sphere,
\begin{equation}\label{eq:sgwb_power_basis_exp}
    \mathcal{P}(\Theta) = \sum_\eta \mathcal{P}_\eta e_\eta (\Theta).
\end{equation}
For a pixel basis, we write
\begin{equation}\label{eq:sgwb_power_pixel_exp}
    \mathcal{P} (\Theta) \equiv \mathcal{P} (\Theta_p) = \mathcal{P}_{p'} \delta(\Theta_p, \Theta_{p'}),
\end{equation}
where $\mathcal{P}_{p'}$ is the power of the signal in each pixel. For a spherical harmonic expansion, 
\begin{equation}\label{eq:sgwb_power_ylm_exp}
    \mathcal{P}(\Theta) = \sum_{l=0}^\infty \sum_{m=-l}^l \mathcal{P}_{lm} Y_{lm}(\Theta),
\end{equation}
where $\mathcal{P}_{lm}$ are the spherical harmonic coefficients of the signal and $Y_{lm}(\Theta)$ are the spherical harmonic basis functions. Note that in general the units of sky power components may be different depending on the basis and conventions used. Here, we assume units of GW sky power are strain power per steradian.

\subsection{Cross-correlation statistic}

The SGWB strain signal is best understood as a superposition of sinusoidal plane waves coming from all directions on the sky,
\begin{equation}\label{eq:plane_wave_exp2}
    h_{\mu\nu} (t, {\bm x}) = \int^{\infty}_{-\infty} \diff f \int_{S^2} \diff \Theta \sum_{P=+, \times} h_P (f, \Theta) e^P_{\mu\nu}(\Theta) e^{i 2\pi f \phi},
\end{equation}
where $\phi = (t - \Theta \cdot \textbf{x} / c)$. %
Here, ${\bm x}$ is a position vector in a general coordinate system. %
A GW detector in location ${\bm x}$ such as an interferometer measures
\begin{equation}
     h(t) = \int_{-\infty}^{+\infty} \diff f \int_{S^2} \diff \Theta \sum_{P = +,\times}F_{P}(f, \Theta)\,h_{P}(f, \Theta)\,e^{i 2\pi f \phi},
    \label{eq:strain_model}
\end{equation}
where $F_P$ is the polarization response function of the detector, defined for example in~\cite{SGWB_review2022_AR}.
As instrumental noise is itself stochastic, this sort of signal is not clearly distinguishable from noise in a single detector, in particular in the case where the signal is weak with respect to the noise and both are hard to model independently.
However, even a weak stochastic background induces a correlated signal across multiple detectors.
In current stochastic searches performed on LIGO--Virgo data, the noise is assumed to be fully independent between detectors, hence the cross-correlation of the data streams yields an optimal statistic for the stochastic signal. The latter is often referred to in the literature as an {\it optimal filter}~\cite{BruceAllen_SGWB_signal}, and we describe its application as an estimator for the SGWB signal in what follows.

Consider the case of a {\it baseline} $I$ made up of two ground-based GW detectors 1, 2 each with data output
\begin{equation}
    s(t) = h(t) + n(t),
\end{equation}
where $h(t)$ denotes the strain due to a SGWB and $n(t)$ denotes the detector noise. %
When detector noise is uncorrelated within the baseline, the expectation value of the cross-correlation between the strain in detector 1, $s_1$, and the strain in detector 2, $s_2$, will be sensitive to the signal component only. This can be intuitively derived as
\begin{align}
    \langle C^I \rangle &= \langle h_1 h_2 \rangle + \langle h_1 n_2 \rangle + \langle h_2 n_1 \rangle + \langle n_1 n_2 \rangle \nonumber \\
    & \simeq \langle h_1 h_2 \rangle + \langle n_1 n_2 \rangle \simeq \langle h_1 h_2 \rangle.
\end{align}
We drop terms $\langle h_1 n_2 \rangle$ and $\langle h_2 n_1 \rangle$ since the GW signals and the instrumental noise are uncorrelated. The angle brackets here refer to an ensemble averaging, which is taken in practice by averaging over time, as well as frequency, baselines, and all other available independent measurements of the signal.

We do not consider correlated noise in our discussion. However, there exists a type of noise, Schumann magnetic resonances caused by the EM field of the Earth, which can mimic a correlated SGWB in the detectors. Several methods have been proposed to mitigate these effects in a GW detector network, including noise subtraction methods~\cite{Coughlin_Schumann_2016, Coughlin_Schumann_2018,PhysRevD.90.023013,PhysRevD.87.123009}, the ``GW Geodesy'' method~\cite{Callister_Schumann_2018, Janssens_Schumann_2022}, and spectral modeling~\cite{PhysRevD.102.102005}.

In practice, it is usually more efficient to work with data divided into smaller time segments and transformed to the frequency domain, making use of the Fast Fourier Transform (FFT)~\cite{FFT_algo} algorithm and parallel processing. In the case we consider here, the data are split into segments of equal duration $\tau$, where $\tau$ is much bigger than the light travel time between the two detectors but small enough so that detector response functions do not change significantly over the interval.
The cross-spectral density (CSD) for a baseline $I$ of two detectors evaluated in time segment labeled $t$ and at frequency $f$ is defined as
\begin{equation}
    C^I(t; f) = \frac{2}{\tau} \Tilde{s}^*_1(t; f) \Tilde{s}_2(t; f) \simeq \frac{2}{\tau} \Tilde{h}^*_1(t; f) \Tilde{h}_2(t; f),
\end{equation}
where $\Tilde{s}(t; f)$ is the short-term Fourier transform (SFT) of $s(t)$ of duration $\tau$. %
For conventions used, please see~\cite{SGWB_Romano_Living_Reviews}. %
Then, by Eq.~\eqref{eq:anisotropic_power_law} and the SFT of Eq.~\eqref{eq:plane_wave_exp2}, the expectation value of $C^I(t; f)$ is given by~\cite{SGWB_Romano_Living_Reviews}
\begin{equation} \label{eq:CSD_integral_form}
    \langle C^I(t; f) \rangle = \tau H(f) \sum_\eta \mathcal{P}_\eta \gamma_\eta^I(t; f),
\end{equation}
where $\gamma_\eta^I(t; f)$ here is the unnormalized overlap reduction function (ORF), which describes the correlated sensitivity of the baseline to the sky and frequency modes of the signal.

In the pixel basis, $\eta\rightarrow p$, so that the unnormalized ORF becomes
\begin{equation}
    \gamma^I_{p;\, tf} = \sum_{P=+,\,\times} F^P_1 (t; \Theta_p) F^P_2 (t; \Theta_p) e^{i 2\pi f \Theta_p \cdot \triangle \mathbf{x}(t) / c},
\end{equation}
where $F^P(t; \Theta_p)$ are detector response functions for $P=\{+, \, \times\}$ plane polarized waves, and $\Theta_p$ is the general direction on the sky discretized in the pixel domain, i.e., it is the direction pointing to the center of the pixel $p$. 
The ORF can be transformed to the spherical harmonic basis by
\begin{equation}
    \gamma^I_{lm;\, tf} = \int_{S^2} \diff \Theta_p \, \gamma^I_{p;\, tf} Y_{lm}^* (\Theta_p).
\end{equation}
Note that the normalization of this function on the whole sky is $5/(8\pi)$~\cite{BruceAllen_SGWB_signal}.

Using compact notation with summation over directions $\Theta$ on the sky implied, we put the \textit{signal model} Eq.~\eqref{eq:CSD_integral_form} in a general basis into matrix form
\begin{equation}\label{eq:CSD_matrix_form}
    \langle C^I_{tf} \rangle = K^I_{t f \eta} \cdot \mathcal{P}_\eta,
\end{equation}
where
\begin{equation}
    K^I_{t f \eta} \equiv \tau H(f) \gamma^I_\eta(t; f). 
\end{equation}

The \textit{noise covariance matrix} for the CSD is subsequently \cite{SGWB_Romano_Living_Reviews}
\begin{align}
    N^I_{tf,t'f'} &\equiv \langle C^I_{tf} C^{I*}_{t'f'} \rangle - \langle C^I_{tf} \rangle \langle C^{I*}_{t'f'} \rangle \nonumber \\
    &\approx \frac{\tau^2}{4} \delta_{tt'} \delta_{ff'} P_{n_1}(t;f) P_{n_2} (t;f),
\end{align}
where the one-sided noise power spectrum $P_n$ satisfies
\begin{align}
    \langle \Tilde{s}(t;f) \Tilde{s}^*(t';f') \rangle &\simeq \langle \Tilde{n}(t;f) \Tilde{n}^*(t';f') \rangle \nonumber \\
    &= \frac{\tau}{2} \delta_{tt'} \delta_{ff'} P_n(t;f).
\end{align}

To lighten the notation in remaining derivation, we drop superscripts for detector baselines and subscripts for function dependencies when there is no confusion.

\subsection{Maximum-likelihood Maps} \label{sec:gwb_max_likelihood}

We assume detector noise is Gaussian and stationary on the timescale $\tau$, and further assume that the SGWB is Gaussian, unpolarized, and its spectral shape $H(f)$ is known\footnote{In case of a non-Gaussian signal, we can expect the approach to be sub-optimal, as the likelihood used does not capture key features of the signal. In case of a polarized background, extra terms to the ORF must be considered~\cite{AR_method}.}.

The likelihood function for the cross-correlation statistic of a single baseline is then (using short-hand notation)
\begin{equation} \label{eq:GWB_likelihood}
    \mathcal{L} (C | \mathcal{P}) \propto \prod_{tf}\exp \bigg[-\frac{1}{2} \chi^2(\mathcal{P}) \bigg],
\end{equation}
where, given the signal model in Eq.~\eqref{eq:CSD_matrix_form}, the chi-squared statistic is
\begin{align}\label{eq:GWB_chi2}
    \chi^2(\mathcal{P}) &\equiv (C - \langle C \rangle)^\dag N^{-1} (C - \langle C \rangle) \nonumber \\
    &= (C - K \cdot \mathcal{P})^\dag N^{-1} (C - K \cdot \mathcal{P}),
\end{align}
where the dot product indicates a sum over spatial indices.

Maximizing the likelihood function Eq.~\eqref{eq:GWB_likelihood} with respect to $\mathcal{P}$ is equivalent to minimizing the chi-squared statistic Eq.~\eqref{eq:GWB_chi2}. By matrix differentiation, we derive the maximum-likelihood estimates of angular power spectrum, the \textit{clean map},
\begin{equation}\label{eq:clean_map}
    \hat{\mathcal{P}}_\eta = \sum_{\eta'}\Gamma^{-1}_{\eta\eta'} X_{\eta'},
\end{equation}
where $X$ is the so-called \textit{dirty map}, and $\Gamma$ is the \textit{Fisher information matrix}. %

The dirty map represents the GW sky seen through the beam matrix of the two detectors and is given by
\begin{equation}\label{eq:dirty_map}
    X_\eta = \sum_{tf} K_{tf\eta}^\dag\, N_{tf}^{-1} \, C_{tf}.
\end{equation}

The Fisher matrix, which can be interpreted as a point spread function, codifying how signals from point sources spread elsewhere due to finite coverage of the sky by a network of GW detectors, is
\begin{equation}\label{eq:fisher_info}
    \Gamma_{\eta \eta'} = \sum_{tf} K^\dag_{tf\eta} N^{-1}_{tf} K_{tf\eta'}.
\end{equation}

The specifics of the derivation are described in~\cite{AR_method}. The clean map statistic can be viewed as a directional extension of the optimal statistic derived in~\cite{BruceAllen_SGWB_signal}, and is thus robust to noise non-stationarity on time-scales longer than the analyzed time segment $\tau$, as it consists of an inverse noise-weighted average over segments.

The above derivation for a baseline of two GW detectors is easily generalized to a multi-detector network. Assuming each baseline provides an independent measurement of the signal, it is sufficient to sum dirty maps and Fisher matrices over all baselines in the network
\begin{equation}
    X = \sum_I X^I, \quad \quad \quad \Gamma = \sum_I \Gamma^I,
\end{equation}
to obtain the network clean map using Eq.~\eqref{eq:clean_map}.

In the weak signal limit, we can further show \cite{SHD_method}
\begin{align}
    \langle X \cdot X^\dag \rangle - \langle X \rangle \langle X^\dag \rangle &\approx \Gamma, \\
    \langle \hat{\mathcal{P}} \cdot \hat{\mathcal{P}}^\dag \rangle - \langle \hat{\mathcal{P}} \rangle \langle \hat{\mathcal{P}}^\dag \rangle &\approx \Gamma^{-1}.
\end{align}
Therefore, $\Gamma$ is the covariance matrix for the dirty map $X$ and $\Gamma^{-1}$ is the covariance matrix for the clean map $\hat{\mathcal{P}}$.

We can then define the signal-to-noise (SNR) map to be the result of the matrix multiplication~\cite{AR_O1}
\begin{equation}\label{eq:snr_map}
    \rho =\Gamma^{-\frac{1}{2}} \cdot \hat{\mathcal{P}},
\end{equation}
which takes off-diagonal elements of the Fisher matrix into account, and the noise standard deviation map to be
\begin{equation}\label{eq:noise_map}
    \sigma = \sqrt{\text{diag} \, \Gamma^{-{1}}}.
\end{equation}
The noise map so defined is only sensitive to diagonal elements of the inverse Fisher matrix, ignoring all pixel-pixel correlations. However, correlations between different locations on the sky are nontrivial. The noise map is thereby only an approximation of the noise standard deviation of the estimator $\hat{\mathcal{P}}$. In the case of a singular Fisher matrix, the calculation of the SNR requires regularizing adjustments as described below.

\begin{figure*}
    \centering
    \subfloat{
        \includegraphics[width=.45\textwidth]{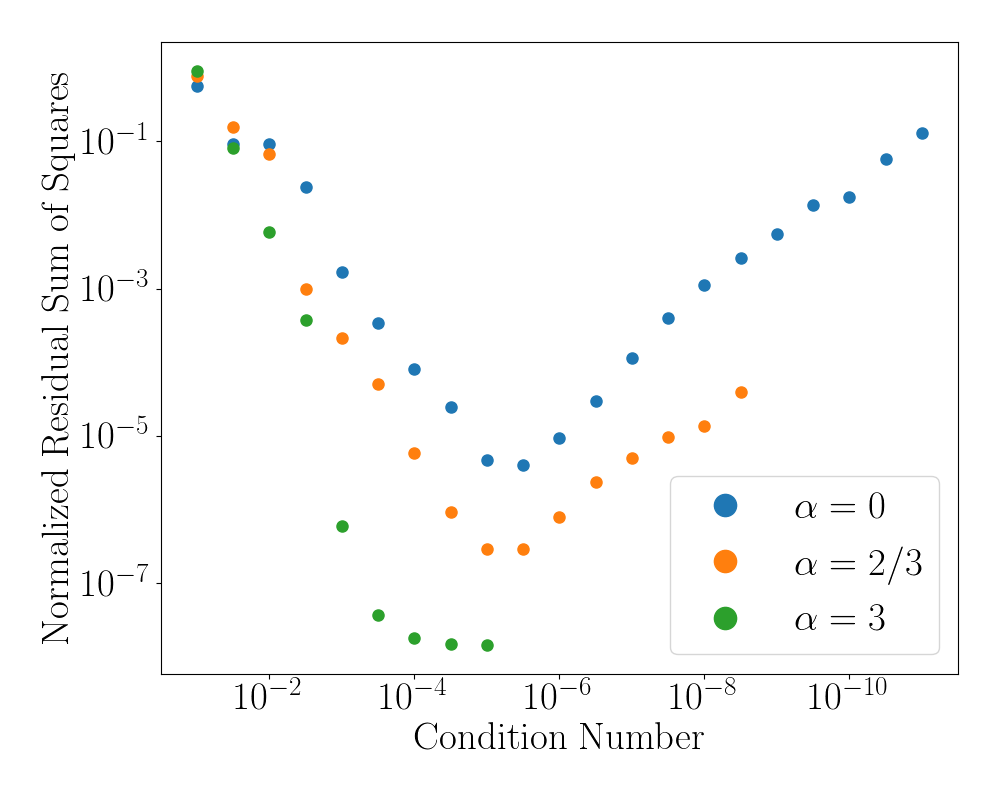}
    }
    \subfloat{
        \includegraphics[width=.45\textwidth]{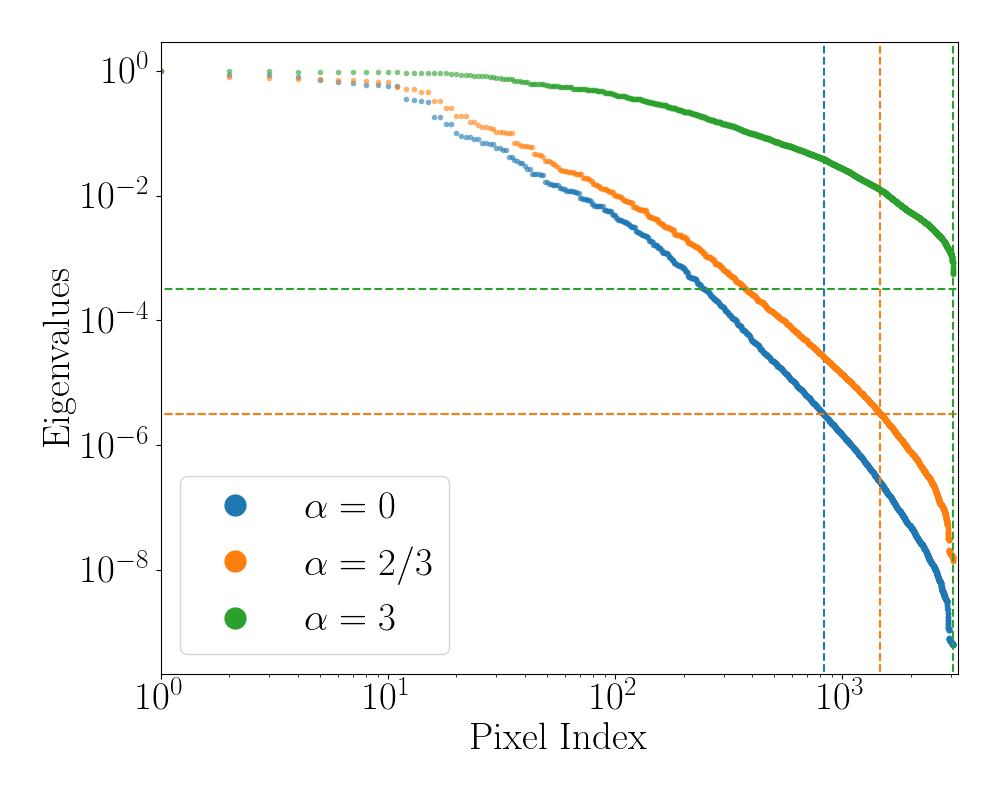}
    }\hfill

    \caption{The left panel shows the condition numbers (i.e., eigenvalue thresholds) and resulting normalized residual sums of squares for power laws of spectral indices 0, $2/3$ and 3. We select the threshold returning the least residual sum of squares in a monopole injected simulation for each spectral index. The right panel illustrates the comparison between the Fisher matrix eigenvalue distributions for the HLV network for different values of $\alpha$. The horizontal dashed lines show the optimal thresholds determined via monopole simulations.}
    \label{fig:condition_num}
\end{figure*}
The dirty maps and Fisher matrices may be calculated over broad frequency bands to improve detection statistics. However, this implies integrating over the spectral shape of the SGWB, $H(f)$, hence the resulting clean map Eq.~\eqref{eq:clean_map} is a biased estimator of the angular power distribution, as we do not know $H(f)$ \textit{a priori}. The standard spectral-model approach is to assume a power-law spectral model $H_\alpha(f)$ as in Eq.~\eqref{eq:spectral_shape} and estimate $\hat{\mathcal{P}}$ for a set of $\alpha$ candidates. We consider here three possible spectral index values, in keeping with the LVK searches \cite{O3_SGWB_anisotropic}: $\alpha=0$, a flat energy density spectrum consistent with many cosmological models \cite{cosmo_SGWB_review}; $\alpha=2/3$, an astrophysical background dominated by CBCs \cite{Regimbau_astro_SGWB}; and $\alpha=3$, a generic flat strain spectrum \cite{SGWB_supernovae}. Other approaches, such as the ASAF approach, solve for $\hat{\mathcal{P}}$ in each frequency bin, and do not require a model for $H(f)$; however, in this case it is not possible to invert the full Fisher matrix, as it is prohibitively singular in a single frequency bin. This is the main motivation for integrating over broader frequency ranges when taking pixel-pixel correlations into account.

\subsection{Deconvolution Regularization} \label{subsec:deconv}

To perform the deconvolution in Eq.~\eqref{eq:clean_map}, we need to invert the Fisher matrix which is typically singular due to the uneven sampling of the sky. In the absence of detections, current search methods employed by both the LVK collaboration and independent groups condition the Fisher matrix in an ad hoc way; specifically, the LVK has proceeded either by restricting only to diagonal elements and hence ignoring all pixel-pixel correlations as in the BBR search for point sources, or discarding the smaller $1/3$ of eigenvalues and fixing a maximum multipole as in the SHD search for extended sources \cite{O3_SGWB_anisotropic}. Other groups have instead chosen a fixed condition number for the Fisher matrix~\cite{AR_O1, AR_O1O2}. It is clear that moving towards the detection era for SGWBs systematic ways to better regularize the Fisher matrix must be established, especially to claim confident detections.

The Fisher matrix is in general singular since there exist null directions the detector network is insensitive to~\cite{SGWB_anisotropy_Allen, AR_method}. 
The current level of singularity may be seen in the right panel of Fig.~\ref{fig:condition_num}, where eigenvalues of the LIGO Hanford--LIGO Livingston--Virgo (HLV) network Fisher matrix with present data are shown. %
To address this issue, we use a singular value decomposition (SVD) \cite{SVD_method} of the Fisher matrix to rank contributing directions and discard eigenmodes which carry little information \cite{SHD_method}. The inherent condition number of the matrix, i.e., the ratio between the minimum and maximum eigenvalues, depends on the spectral shape. Including too many eigenmodes introduces unwanted noise whereas discarding too many eigenmodes sacrifices signals. The SVD technique allows to condition the matrix, i.e., impose an eigenvalue threshold such that all normalized eigenvalues that are smaller than the imposed condition number are discarded. This approach was previously explored systematically in~\cite{Deepali_fisher}. In the rest of this paper, the notation $\Gamma^{-1}$ indicates a regularized inverted Fisher matrix, and $\Gamma$ is the regularized Fisher matrix. In particular, in calculating the SNR as presented in Eq.~\eqref{eq:snr_map}, we employ the square root of the regularized Fisher matrix.

We present a method to determine this threshold empirically via simulations. For each spectral index $\alpha$, we impose the condition number returning the least residual sum of squares (RSS) from a respective high SNR monopole simulation, 
\begin{equation}
    \mathrm{RSS} = (\mathcal{P}_\mathrm{inj} - \hat{\mathcal{P}})^T (\mathcal{P}_\mathrm{inj} - \hat{\mathcal{P}}),
\end{equation}
where $\mathcal{P}_\mathrm{inj}$ is the injected monopole. Monopole simulations are used for the calculation since we expect an intrinsic monopole irrespective of spectral shapes, and stronger than any higher multipoles \cite{SGWB_review2022_AR}. The monopole simulations are generated using the HLV detector network configuration in its O3 sensitivity, since most sensitivity of the combined O1+O2+O3 runs comes from O3. The residuals computed for different condition numbers are illustrated in the left panel of Fig.~\ref{fig:condition_num}. The optimal condition numbers with the smallest residuals for different power laws are listed in Table \ref{table:condition_num}. The percentages of eigenvalues kept using the optimal condition numbers are also shown in Table \ref{table:condition_num}. Note these are quite different from the nominal value of $2/3$ in the LVK SHD searches, and depend strongly on the spectral index. The comparison between the Fisher matrix eigenvalues and the associated optimal condition numbers for the HLV network for different power laws is shown in the right panel of Fig.~\ref{fig:condition_num}.

\begin{table}
\begin{center}
\begin{tabular}{>{\centering\arraybackslash}p{1cm} | >{\centering\arraybackslash}p{3cm} > {\centering\arraybackslash}p{3cm}}
  \multicolumn{1}{c|}{$\alpha$} & \multicolumn{1}{c}{Condition number} &  \multicolumn{1}{c}{Percentage of eigenvalues} \\
 \hline
 0 & $10^{-5.5}$ & 27.51\% \\ 
 $2/3$ & $10^{-5.5}$ & 48.93\% \\ 
 $3$ & $10^{-3.5}$ & 100\% \\
 \hline
\end{tabular}
\end{center}
\caption{\label{table:condition_num}Optimal condition numbers and associated percentages of eigenvalues kept for Fisher matrices of the HLV network in its O3 sensitivity. Results are computed empirically via monopole injected simulations in a pixel basis of $N_\mathrm{pix}=3072$ pixels.}
\end{table}

A GW detector network is diffraction-limited, i.e., the resolution and hence the point spread function inherently depends on the frequency of the source. Choosing a pixel basis with a higher resolution than the internal detector resolution at the relevant signal frequencies compromises SNRs of the deconvolved map. The angular resolution of a two-detector baseline is estimated by the diffraction limit~\cite{diffraction_limit}
\begin{equation}\label{eq:diffraction_limit}
    \triangle \theta \simeq \frac{\lambda}{2D} = \frac{c}{2fD},
\end{equation}
where $D$ is the baseline length. The expected angular resolution $\ell_{\rm max} = \pi/\Delta\theta$ per frequency for our analyses can be derived from Eq.~\eqref{eq:diffraction_limit}, with some technicalities.
The HL baseline length $D_\mathrm{HL} = 3002$ km is used throughout the analyses for being the most sensitive baseline, so we expect this baseline to dominate the resolution.
Furthermore, for broadband analyses we expect each frequency to contribute as a function of overall signal spectral shape~\cite{O3_SGWB_anisotropic}.
While we quote results at a fixed reference frequency in this case, chosen in line with previous results, this frequency does not determine our angular resolution.
On the other hand, for the banded approach described below, we quote results at the midpoint of each frequency band considered. %
As these are not broad-band integrated results, the reference frequency used here can give a reasonable estimate of the expected angular resolution in each band. 
Note that a recent study shows that the diffraction limit is not optimal to resolve sources~\cite{Floden22}, however we are most concerned here with maximizing the detection statistic, not the recovered resolution.
In our pixel-basis approach, we use this limit as a lower bound on the number of pixels to use, so as to over-resolve the anisotropies. The upper bound on pixel number is set by Fisher matrix regularization, as described below.

Finally, adding more detectors to the network is a form of regularization, since it provides larger coverage of the sky. With ever-improving sensitivities of existing detectors and addition of new detectors (KAGRA~\cite{KAGRA_det} and LIGO India~\cite{LIGO_India}) in the future, Fisher matrices in modeled broadband searches will be much better conditioned so that specialized regularization techniques will become less important. On the other hand, however, the spectral-model-independent method described in Sec.~\ref{sec:gwb_model_indep} relies heavily on trustworthy regularization of all Fisher matrices in its narrowband searches.


\subsection{Multipole Moments}

Extended anisotropies are parameterized in multipole moments of the power on the sky, which are quantified by their spherical harmonic coefficients. We carry out our analysis in the pixel domain by choosing a pixel basis expansion as in Eq.~\eqref{eq:sgwb_power_pixel_exp}. Hence, to obtain limits on anisotropies about the mean background, we convert from the pixel basis to the spherical harmonic basis.

We can construct estimators of spherical harmonic coefficients $\hat{\mathcal{P}}_{lm}$ for the GW sky directly using estimated angular power in pixels $\hat{\mathcal{P}}_p$ by
\begin{equation}
    \hat{\mathcal{P}}_{lm} = (Y^\dag \cdot \Gamma \cdot Y)^{-1} \cdot (Y^\dag \cdot \Gamma \cdot \hat{\mathcal{P}}_p),
    \label{eq:Plms}
\end{equation}
where $Y=Y_{lm,p}$ is the spherical harmonic basis matrix. Noise in the Fourier domain can be computed as~\cite{AR_O1O2}
\begin{equation}
    \hat{N}_l = \frac{1}{1+2l} \sum_m \bigg| \sum_{pp'} Y_{lm,p} \Gamma_{pp'} Y^*_{p',lm} \bigg|^2.
\end{equation}
Analogous to the approach in CMB experiments, we construct unbiased estimators of the squared angular power $\hat{C}_l$ in the spherical harmonic basis by
\begin{equation}\label{eq:sgwb_cls}
    \hat{C}_l = \frac{1}{1+2l} \sum_m |\hat{\mathcal{P}}_{lm}|^2 - \hat{N}_l.
\end{equation}

Assuming a spectral index $\alpha$, our maximum-likelihood estimates $\hat{\mathcal{P}}_\eta$ of the GW angular power spectrum yield an estimate of the normalized GW energy density $\Omega_\mathrm{GW}$ at a reference frequency $f_\mathrm{ref}$, integrated over a broad band of frequencies. The normalized GW energy density at the reference frequency $f_\mathrm{ref}$ is calculated using the noise-weighted monopole value $ \hat{\mathcal{P}}_{00}$ of the GW power across the sky estimated from the maps by Eq.~\eqref{eq:Plms},
\begin{equation}\label{eq:omega_upper_lim}
    \Omega_\alpha \equiv \Omega_\mathrm{GW} (f_\mathrm{ref}) = \frac{2 \pi^2}{3 H_0^2} f_\mathrm{ref}^3 \hat{\mathcal{P}}_{00}.
\end{equation}
Note the computation of $\hat{\mathcal{P}}_{00}$ includes a normalization by a factor of $5/(8\pi)$ due to the normalization of detector overlap functions~\cite{BruceAllen_SGWB_signal}. The GW energy density spectrum at arbitrary frequencies is then obtained by re-scaling the frequency-integrated estimate of $\Omega_\mathrm{GW}$ with its spectral shape,
\begin{equation}
    \Omega_\mathrm{GW}(f)= \Omega_\alpha \bigg( \frac{f}{f_\mathrm{ref}} \bigg)^\alpha.
\end{equation}

%% file: sections/3_unmodeled_method.tex
\section{Spectral-Model-independent Approach} \label{sec:gwb_model_indep}

In Eq.~\eqref{eq:anisotropic_power_law}, we assume the GW power on the sky can be factored into separate directional and frequency components, and we further assume the spectral shape is a power law of index $\alpha$ as in Eq.~\eqref{eq:spectral_shape}. Though these two simplifications are motivated by many astrophysical and cosmological models \cite{Regimbau_astro_SGWB, cosmo_SGWB_review}, they are not exact and will eventually break down. 

There are SGWBs with non power-law spectral shapes. For example, in low frequencies, the SGWB due to CBCs is well modeled by a power law of spectral index $2/3$. However, in high frequencies, we expect a spectral turnover determined by the redshift-dependent star formation rate and the average total mass of binary black holes (BBHs)~\cite{turnover1, turnover2}. Measuring this turnover will thus allow us to probe the average BBH total mass, the evolution of that quantity over cosmic time, and the star formation history of the Universe. Moreover, there may even be backgrounds with direction-dependent spectral emission, which the spectral-model search is not optimal for. 

A generic, spectral-model-independent approach thus allows us to probe the spectral shape of the SGWB and potentially identify contributing sources and mechanisms. Towards building a general, model-agnostic search for SGWBs, a first step is to reduce the assumption of spectral shapes to a minimum while maintaining the GW strain power factorization.

\subsection{Adaptive Frequency Banding}

To reconstruct the spectral dependence of a SGWB, we run map-making in distinct frequency bands of adaptively chosen bandwidths. The number of bands is a user input to the pipeline, which ideally is numerous enough to achieve a good approximation of the spectral shape. Nevertheless, it competes with the conditioning of the Fisher matrix in each band. Each band needs to be wide enough for the Fisher matrix to be adequately well-behaved so as to allow inversion. With the number of bands as input, the algorithm chooses frequency endpoints with each band containing equal amount of noise-weighted strain power. Within each band, we then assume a fixed, least-informative prior flat in energy density, $H_\alpha(f) \sim f^{-3}$, to run map-making. The optimal condition number of the Fisher matrix for each band would ideally be determined independently using the method presented in Sec. \ref{subsec:deconv} with a monopole simulation in that band. However, running a simulation for each band is computationally expensive. At the current sensitivity level, we choose to use the broadband optimal condition number in Table \ref{table:condition_num} for each spectral shape as a proxy. With the assumption of an angular-independent spectral shape and with the Fisher matrix properly conditioned, estimated GW energy densities in each frequency band trace out the strain power spectral dependence. 

\subsection{Adaptive Pixelization}

In the spectral-model-independent method, a single angular resolution does not accommodate all frequency bands due to different diffraction limits estimated via Eq.~\eqref{eq:diffraction_limit}. Fixing an angular resolution across all bands over-resolves lower frequencies and hence impairs the Fisher matrix conditioning, and under-resolves higher frequencies and hence loses attainable SNRs. We therefore independently estimate the expected angular resolution for each frequency band using Eq.~\eqref{eq:diffraction_limit}, with $D=D_\mathrm{HL}$ and $f$ to be the midpoint of the band. %
We limit ourselves to the optimal resolutions within the \texttt{HEALPix} package~\cite{healpix} (i.e., choices of resolution $N_{\rm side} = 2^n$), implying a coarse resolution variation over frequency bands, as described in Sec.~\ref{sec:simulations}. %
We choose the pixel resolution such that point sources are over-resolved: specifically, spread across four pixels. %
We leave the pixelization optimization as well as the exploration of alternative pixelization schemes which allow for finer resolution variations for future work.

%% file: sections/4_simulations.tex
\section{Simulations} \label{sec:simulations}

\begin{figure*}
    \centering
    \subfloat{
        \includegraphics[width=.28\textwidth]{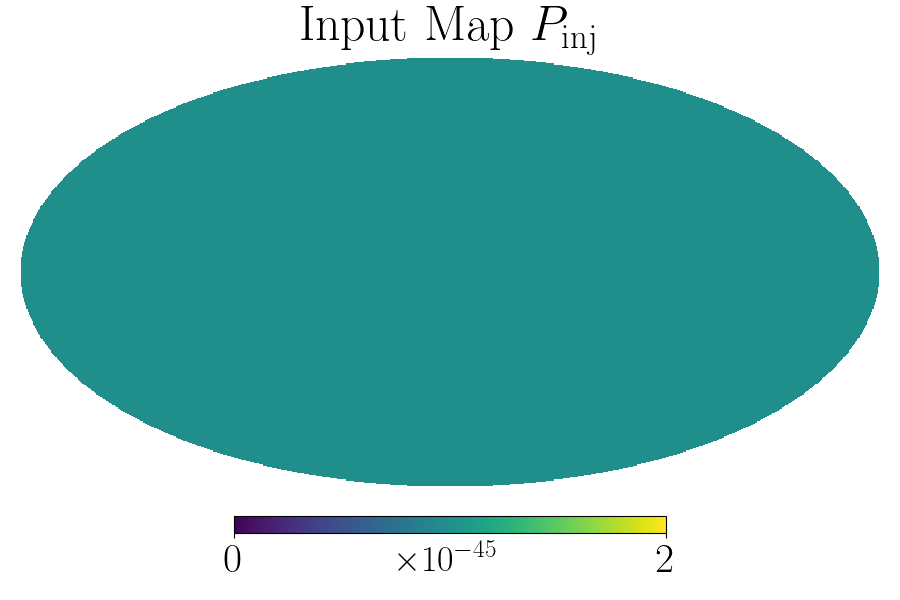}
    }
    \subfloat{
        \includegraphics[width=.28\textwidth]{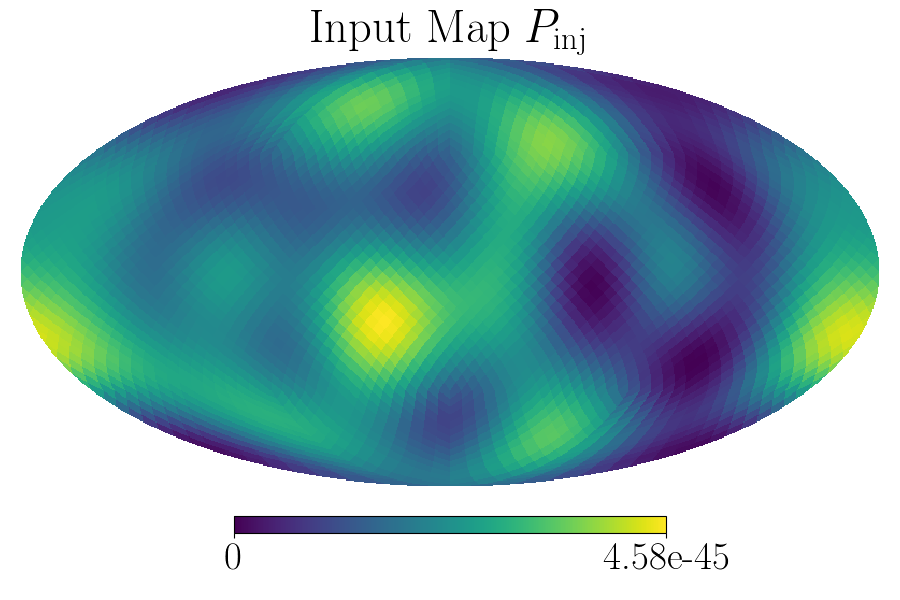}
    }
    \subfloat{
        \includegraphics[width=.28\textwidth]{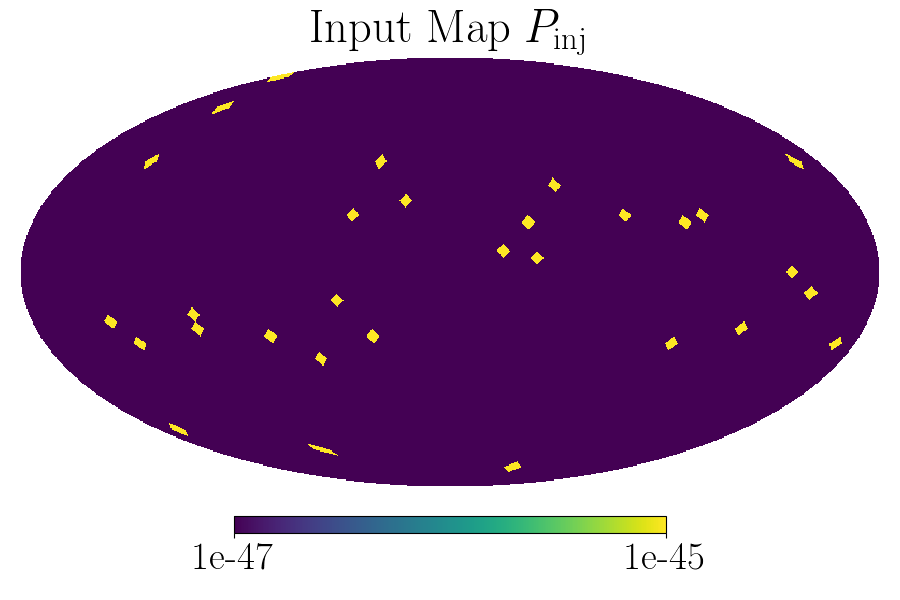}
    }

    \subfloat{
        \includegraphics[width=.28\textwidth]{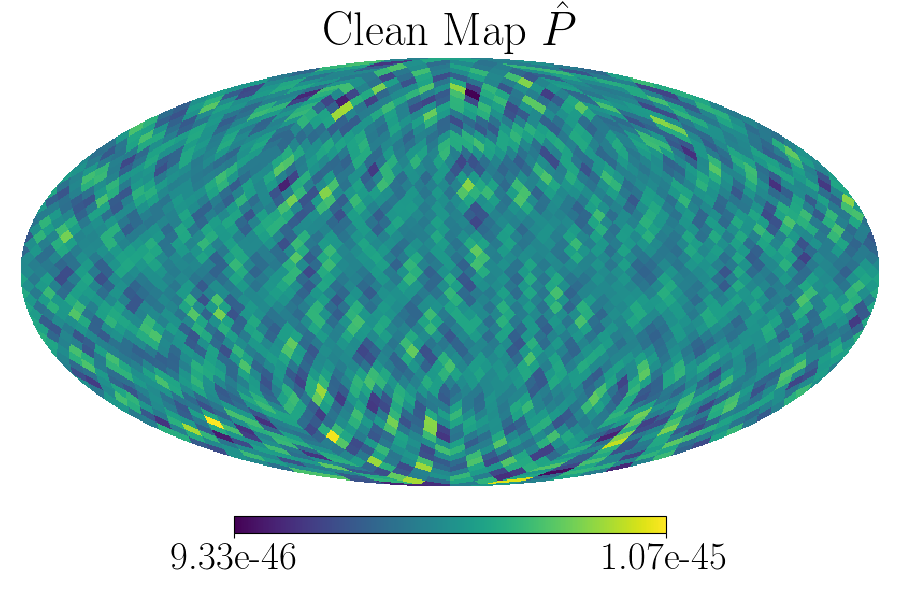}
    }
    \subfloat{
        \includegraphics[width=.28\textwidth]{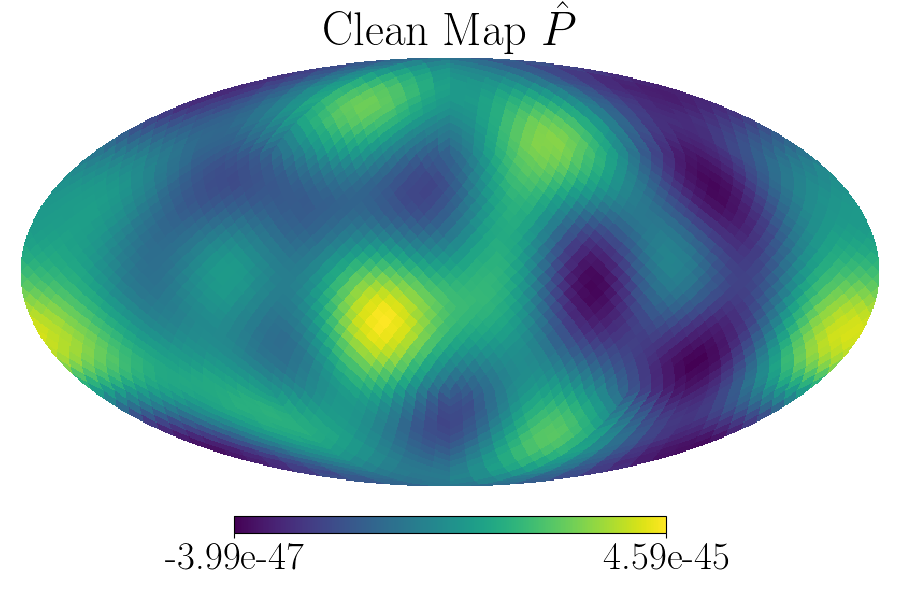}
    }
    \subfloat{
        \includegraphics[width=.28\textwidth]{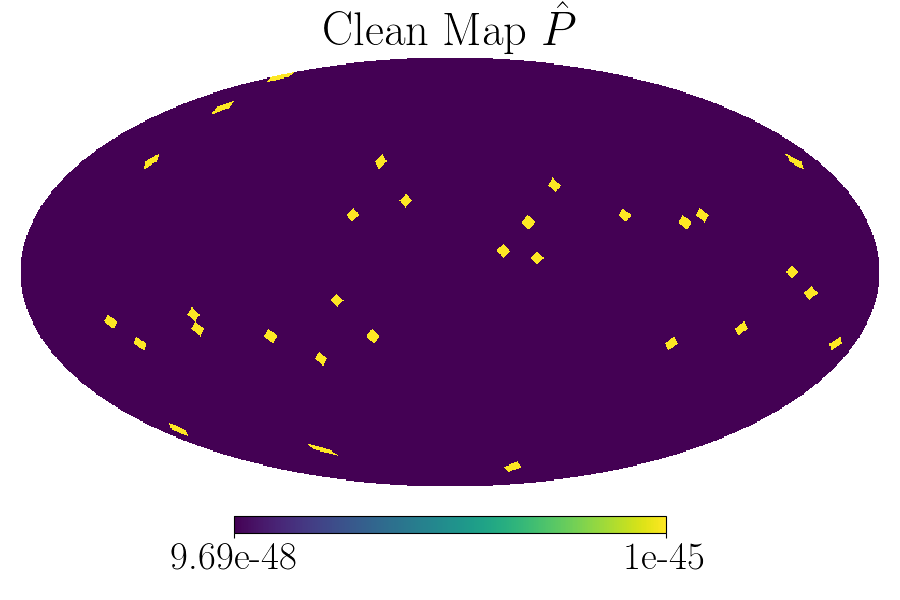}
    }

    \subfloat{
        \includegraphics[width=.28\textwidth]{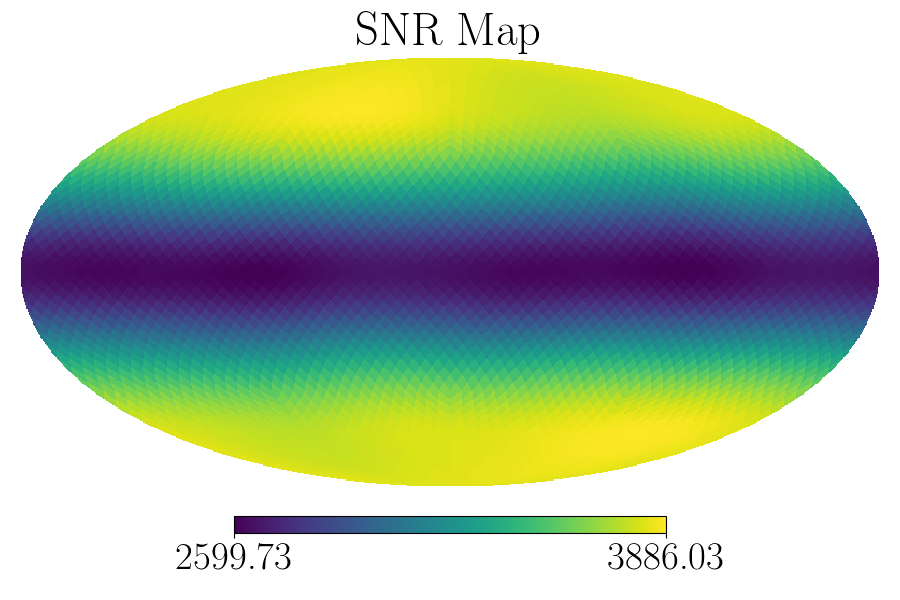}
    }
    \subfloat{
        \includegraphics[width=.28\textwidth]{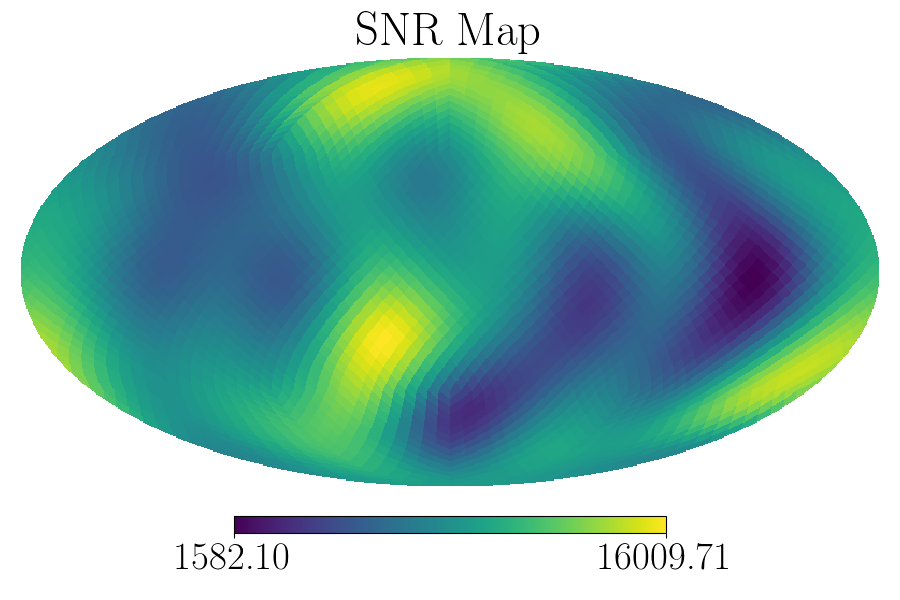}
    }
    \subfloat{
        \includegraphics[width=.28\textwidth]{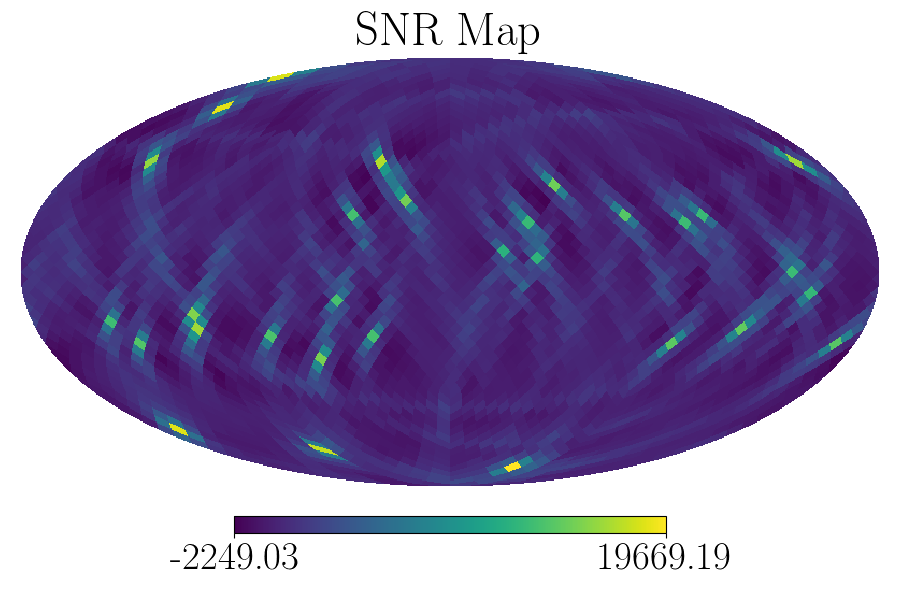}
    }

    \subfloat{
        \includegraphics[width=.28\textwidth]{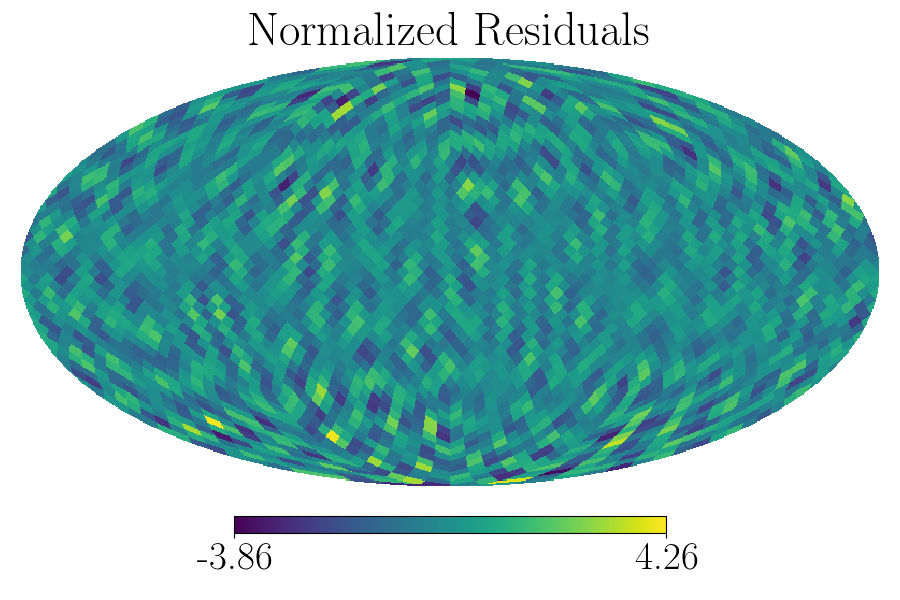}
    }
    \subfloat{
        \includegraphics[width=.28\textwidth]{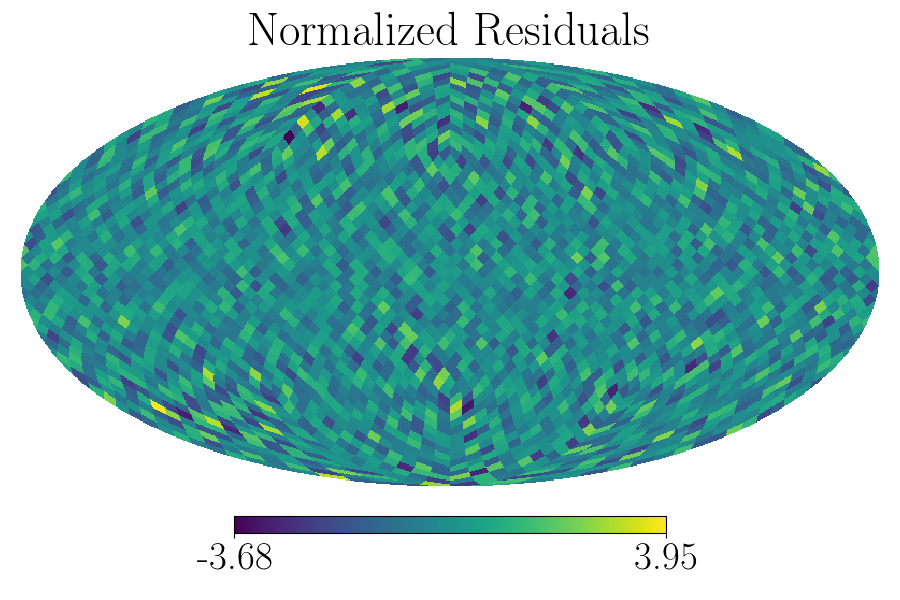}
    }
    \subfloat{
        \includegraphics[width=.28\textwidth]{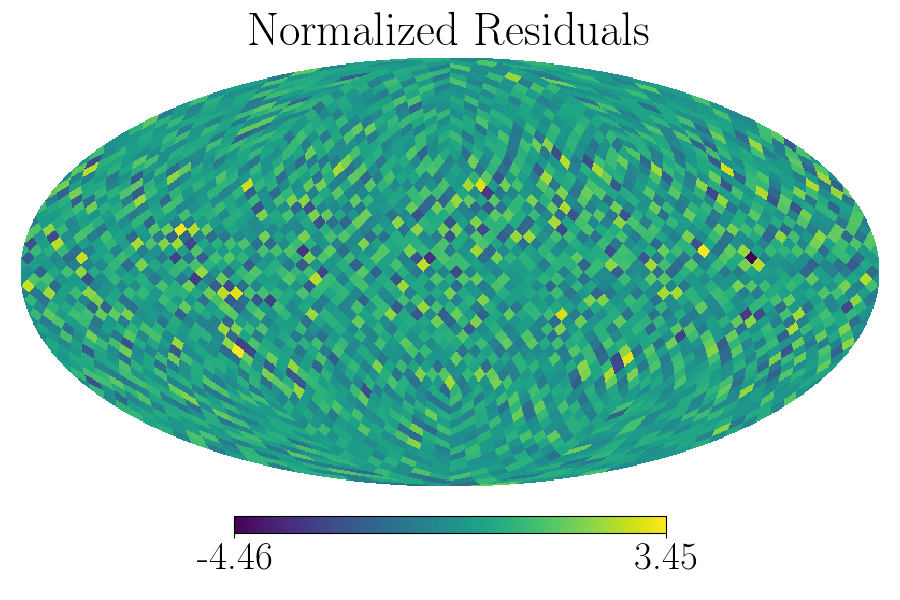}
    }
    
    
    \caption{Top to bottom: input, clean, SNR, and normalized residual maps from simulations described in Sec.~\ref{sec:simulations}. From left to right: monopole ($\alpha=0$), Gaussian random field ($\alpha=2/3$), 30 random points ($\alpha=3$). %
    For all simulations, the pixel with the maximum residual is at the level of a few percent of the injected signal. We have verified that the residuals are Gaussian distributed with norm 1. %
    In the monopole reconstruction, the SNR map presents a characteristic horizontal band due to the shape the ORF traces on the sky over 1 day; this may be also noticed in Fig.~\ref{fig:narrowband_maps}. %
    In the Gaussian field case, the injected map has patches of zero power, and is thus more subject to poor estimation due to noise fluctuations than the other cases shown. %
    In the case of 30 random points, 
    the SNR map presents a residual of the point-spread function with negative values as it is the result of the matrix operation in Eq.~\ref{eq:snr_map}, which can give rise to negative fluctuations where the pixel power is very low.
    }
    \label{fig:sim_mapping}
\end{figure*}
\begin{figure*}
    \centering
    \subfloat{
        \includegraphics[width=.28\textwidth]{figures/O3_sim_nside16_mid_randpoints_HLV_input_map_20-1726Hz.png}
    }
    \subfloat{
        \includegraphics[width=.28\textwidth]{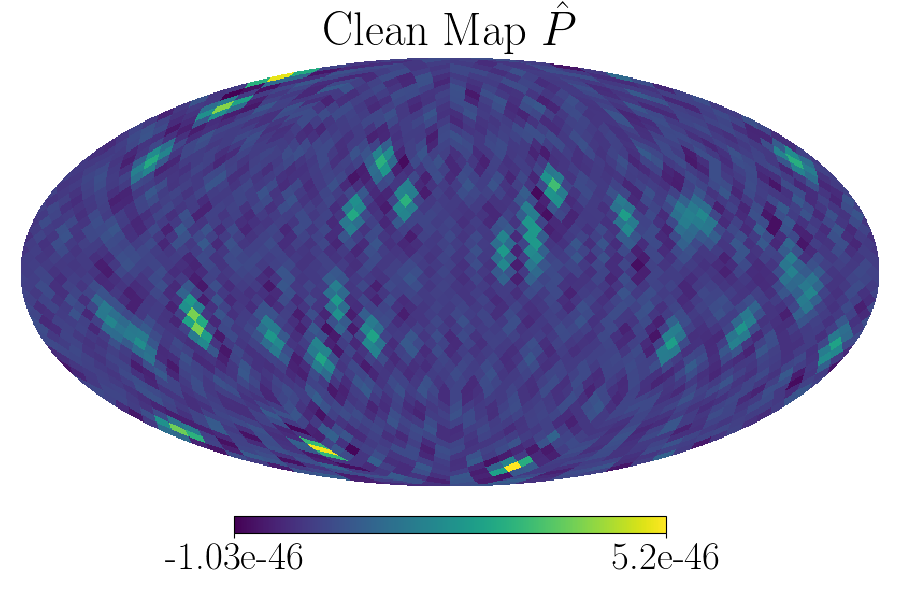}
    }
    \subfloat{
        \includegraphics[width=.28\textwidth]{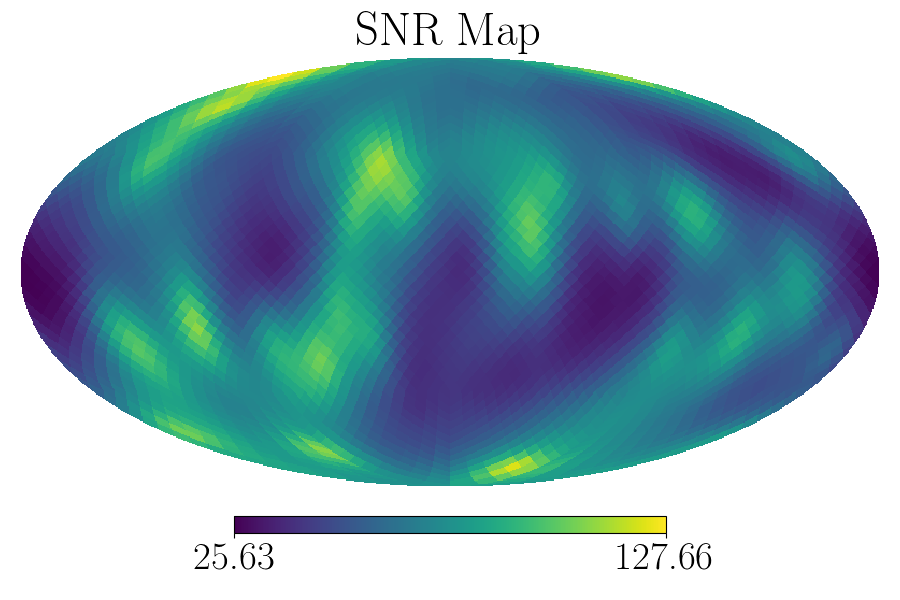}
    }
     
    \caption{Example of recovery of 30 point sources for $\alpha=0$. The smearing of the point spread is not completely deconvolved due to the singularity of the Fisher matrix in this case. The negative power in $\hat P$ is due to noise fluctuations.
    }
    \label{fig:randpoints_alpha0}
\end{figure*}

\begin{figure*}
    \subfloat{
        \includegraphics[width=.42\textwidth]{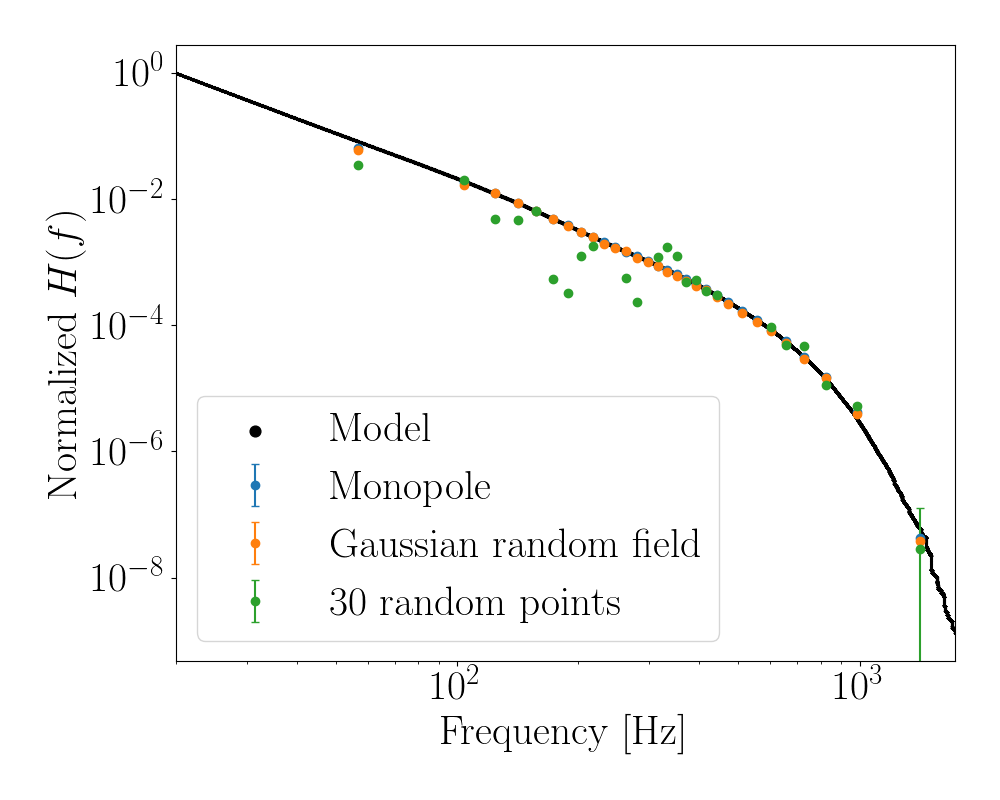}
    }
    \subfloat{
        \includegraphics[width=.42\textwidth]{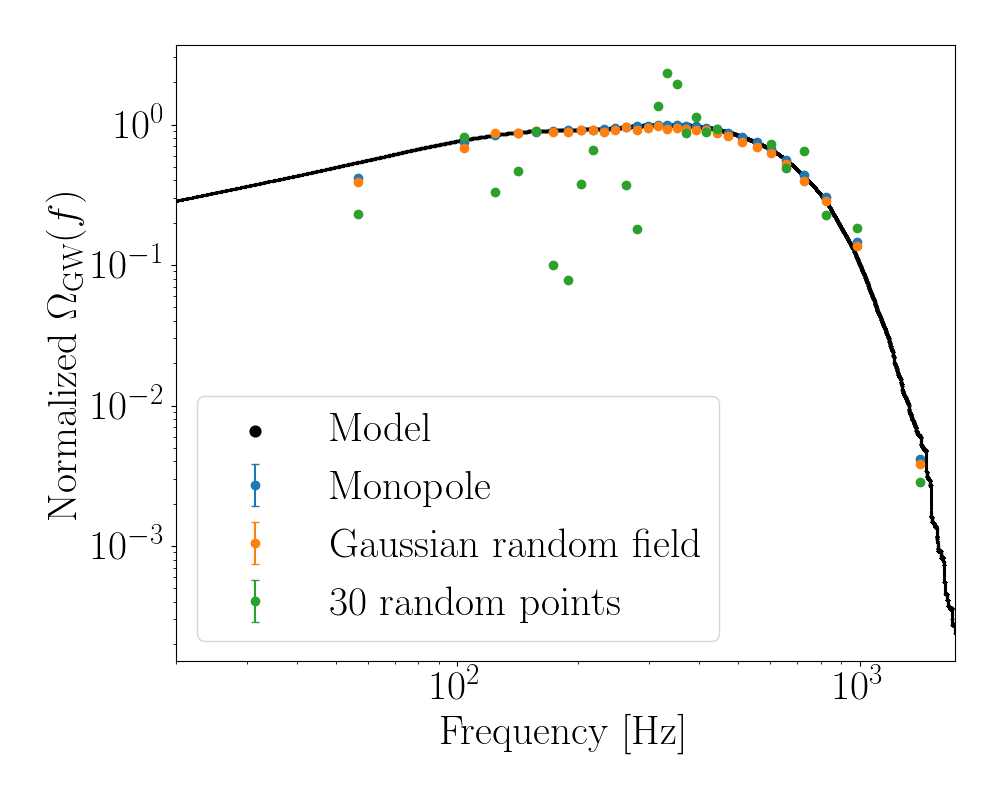}
    }
     
    \caption{Reconstructed normalized energy densities and spectral dependencies in 30 adaptive frequency bands for maps of a monopole (blue), a Gaussian random field (orange), and 30 random point sources (green) for a non-power-law spectral shape. On the left: $H(f)$; on the right: $\Omega(f)$, as defined in Eqs.~\eqref{eq:anisotropic_omega_strain} and~\eqref{eq:anisotropic_power_law}. The spectral shapes are well reconstructed for maps of extended sources (note that the recoveries almost overlap in the plots), whereas for point sources reconstruction is imperfect in particular in the lower frequency bands. See the text for details.}
    \label{fig:sim_reconst_shape}
\end{figure*}

\begin{figure*}
    \subfloat{
        \includegraphics[width=.3\textwidth]{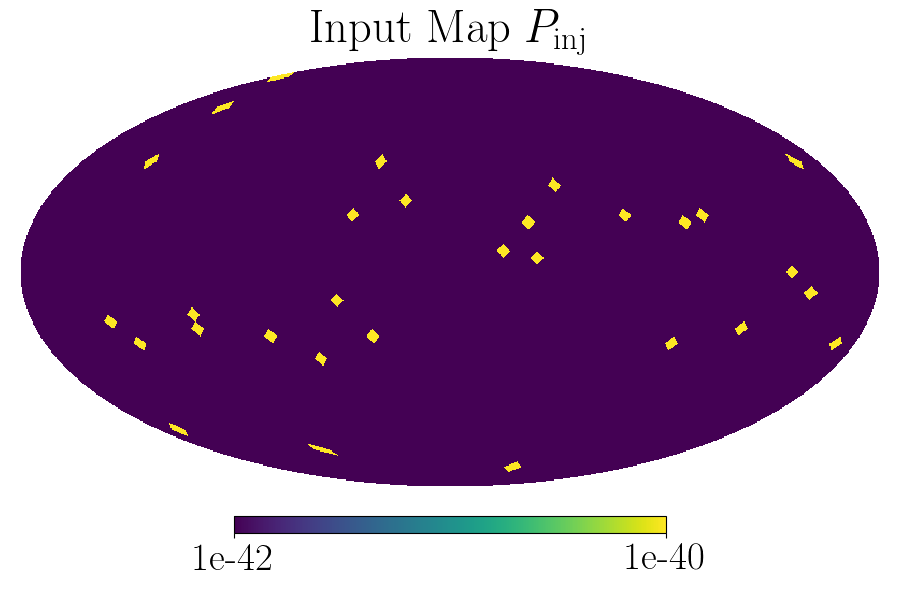}
    }
    \subfloat{
        \includegraphics[width=.3\textwidth]{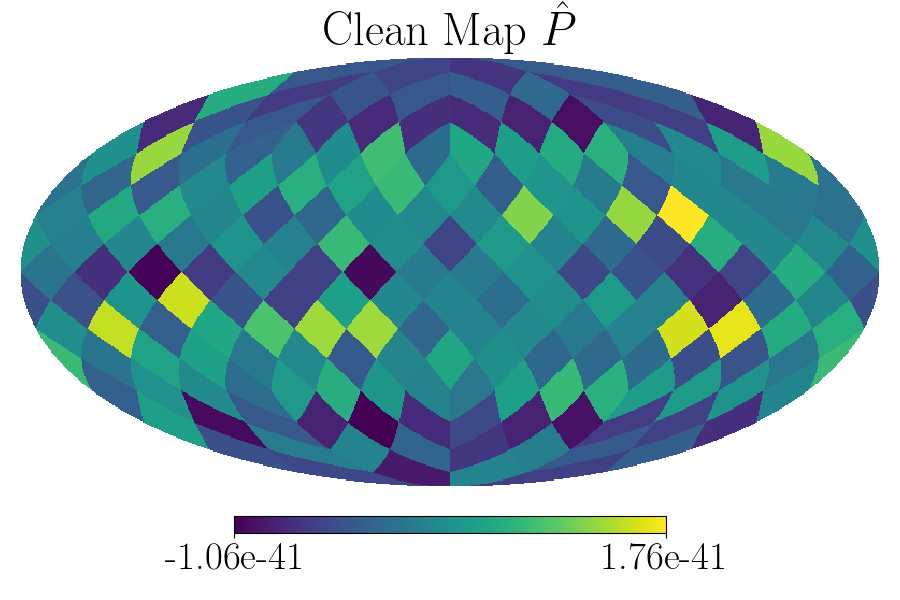}
    }
    \subfloat{
        \includegraphics[width=.3\textwidth]{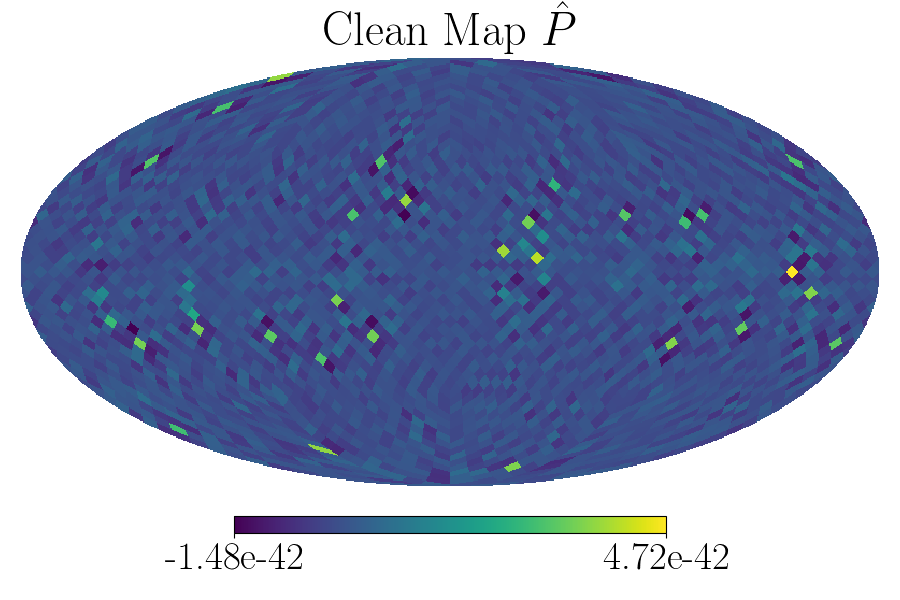}
    }
     
    \caption{Left to right are input map for the spectral-model-independent run in 30 adaptive frequency bands and reconstructed clean maps for bands [20, 93.46875] Hz and [1092.71875, 1726] Hz.}
    \label{fig:sim_randpoint_maps}
\end{figure*}

We demonstrate the maximum-likelihood mapping method in the pixel domain outlined in Sec.~\ref{sec:gwb_max_likelihood} in the spectral-model-dependent case by running our pipeline to recover various injected maps for power-law models of spectral indices 0, $2/3$ and 3. We also illustrate the spectral-model-independent approach to probe spectral dependence as described in Sec.~\ref{sec:gwb_model_indep} via simulations using a realistic spectral shape from the population studies of GWTC-3 \cite{gwtc3_pop}. We use the present sensitivity from the HLV detectors to construct the simulations, released publicly in~\cite{folded_data_O1-O3}. For all spectral models, we inject loud angular power distributions of monopoles, Gaussian random fields, and random point sources on the sky. The simulated input strain power is $h^2 \sim \mathcal{O}(10^{-45})$ for the map-making verification, while we use $h^2 \sim \mathcal{O}(10^{-40})$ for the spectral-model-independent method testing. These may be considered very high SNR cases at present sensitivity, as may be observed in the SNR maps presented in Fig.~\ref{fig:sim_mapping}.

The simulated data consist of sequential CSD frequency segments corresponding to time segments of $\tau = 96$ s over a sidereal day, same as the LIGO--Virgo folded dataset format~\cite{folded_data_O1-O3}, where changes in ORFs are negligible and the noise within each segment stays constant to a good approximation. We simulate CSDs for all three baselines in the HLV detector network in [20, 1726] Hz in both the spectral-model-dependent and independent cases. We then run the complete analysis pipeline to compute maximum-likelihood map solutions for comparisons with injected maps.

We generate simulated CSD time series via Eq.~\eqref{eq:CSD_matrix_form}. We also add simulated Gaussian noise consistent with the representative LIGO noise curve~\cite{H1_sensitivity, L1_sensitivity}
. To verify the pipeline implementation, we use Eqs.~\eqref{eq:clean_map} -- \eqref{eq:fisher_info} to compute $\hat{\mathcal{P}}$. 

With expected SGWBs and associated spectral indices in mind, we demonstrate the map-making functionality in three cases: a monopole map with $\alpha=0$; a Gaussian random field with $\alpha=2/3$ and an maximum resolution $\ell_{\rm max}=8$; and a map of 30 random point sources with $\alpha=3$. We run our searches in the frequency range of [20, 1726] Hz, similarly to the LVK broadband searches in O3 \cite{O3_SGWB_anisotropic}. We choose a pixel basis of 3072 pixels, or equivalently $N_\mathrm{side}=16$ in the \texttt{HEALPix} scheme, with each pixel covering 13.4 deg$^2$.

\begin{figure*}
    \centering
    \subfloat{
        \includegraphics[width=.29\textwidth]{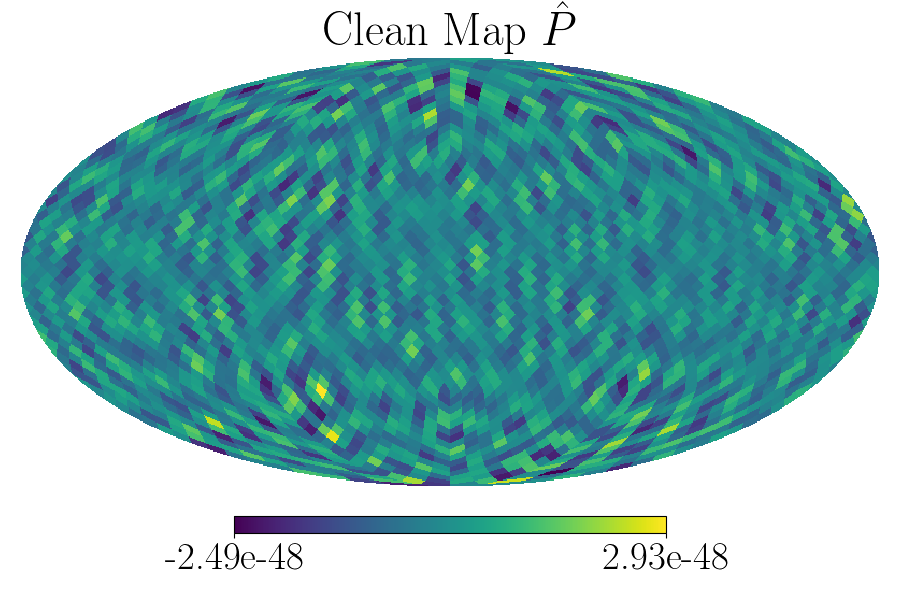}
    }
    \subfloat{
        \includegraphics[width=.29\textwidth]{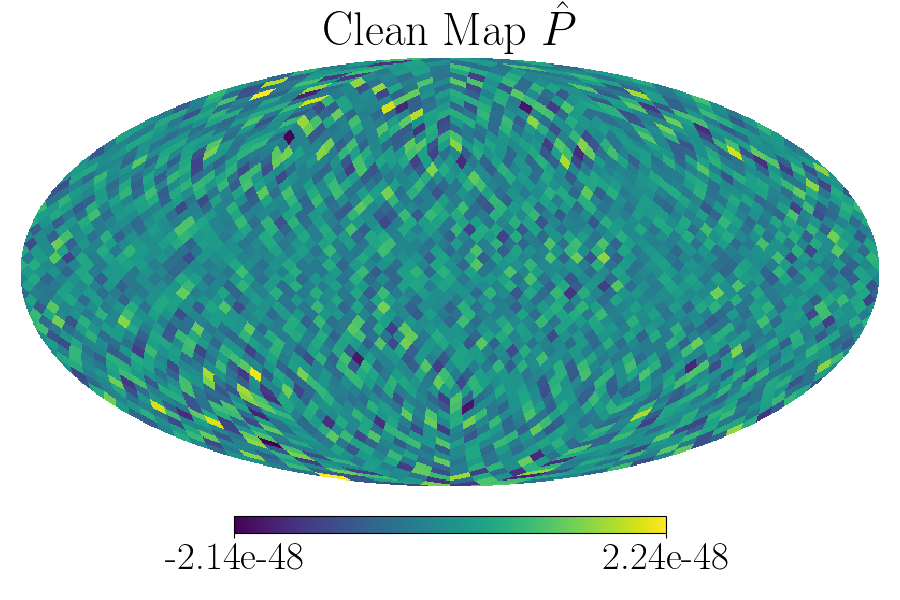}
    }
    \subfloat{
        \includegraphics[width=.29\textwidth]{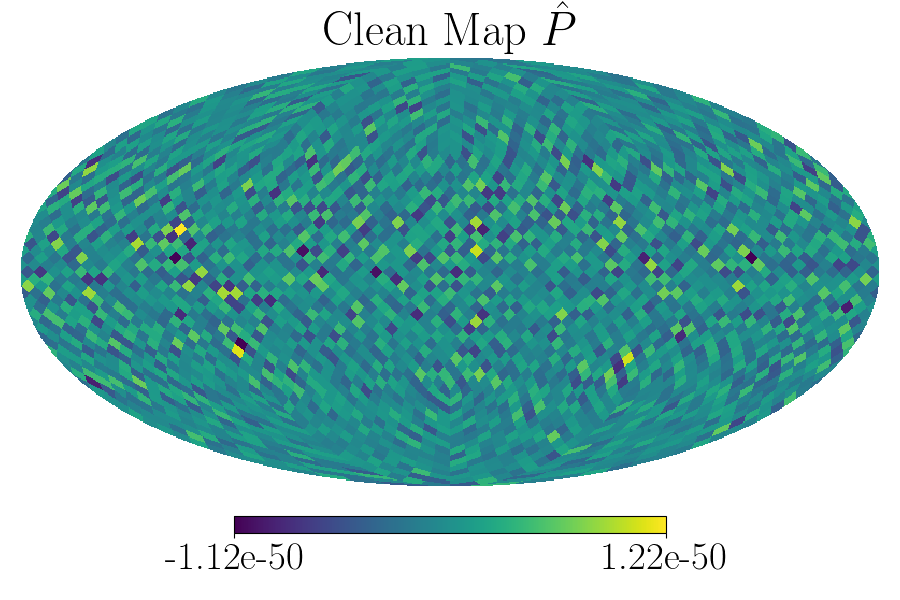}
    }

    \subfloat{
        \includegraphics[width=.29\textwidth]{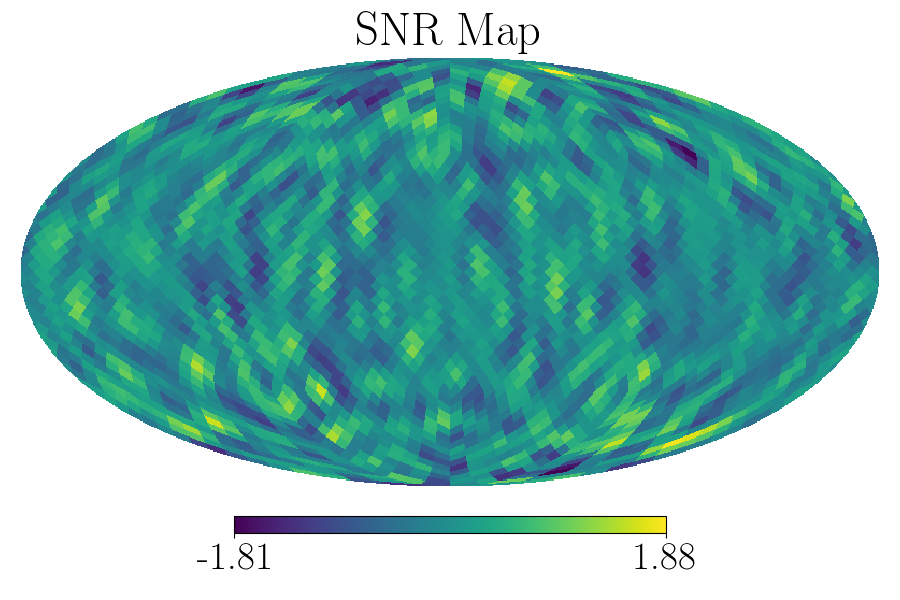}
    }
    \subfloat{
        \includegraphics[width=.29\textwidth]{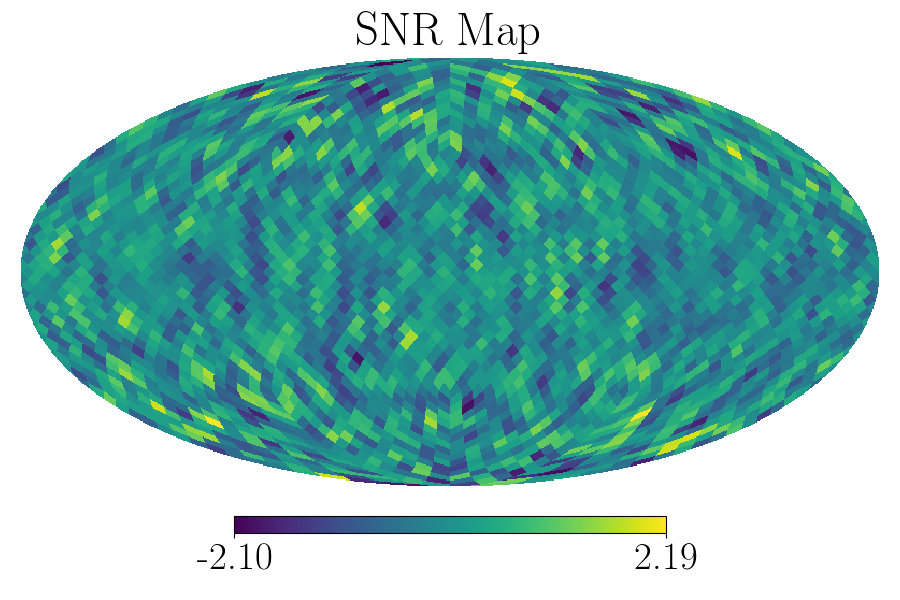}
    }
    \subfloat{
        \includegraphics[width=.29\textwidth]{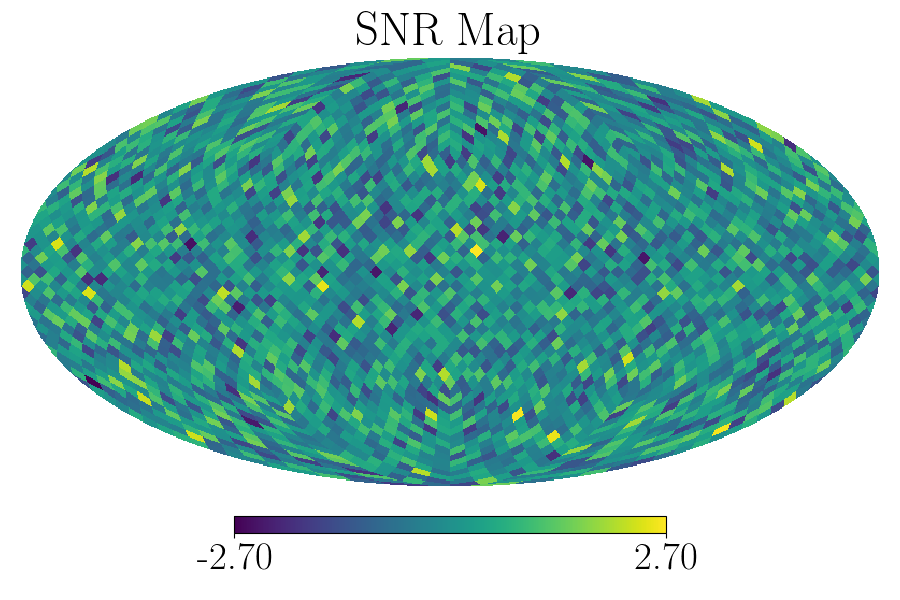}
    }

    \subfloat{
        \includegraphics[width=.29\textwidth]{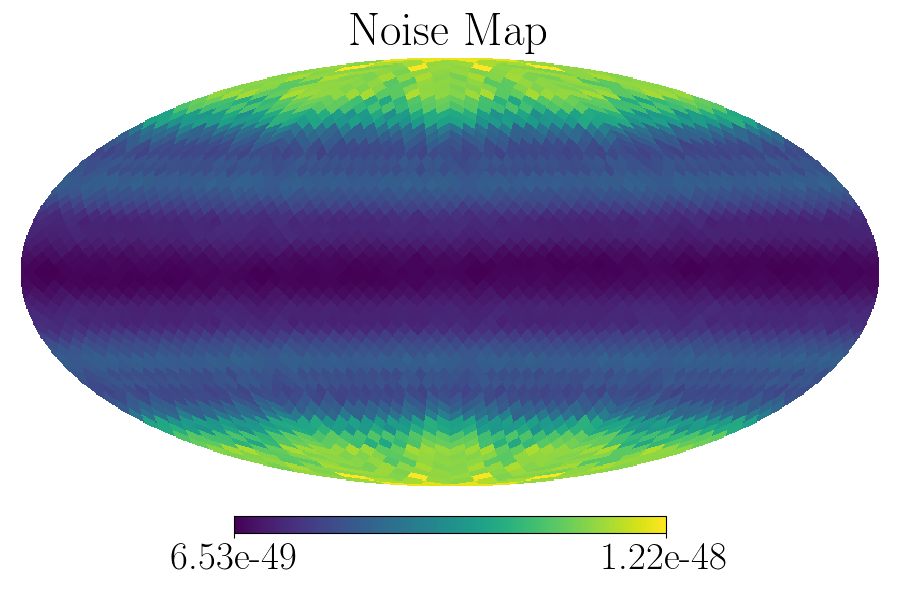}
    }
    \subfloat{
        \includegraphics[width=.29\textwidth]{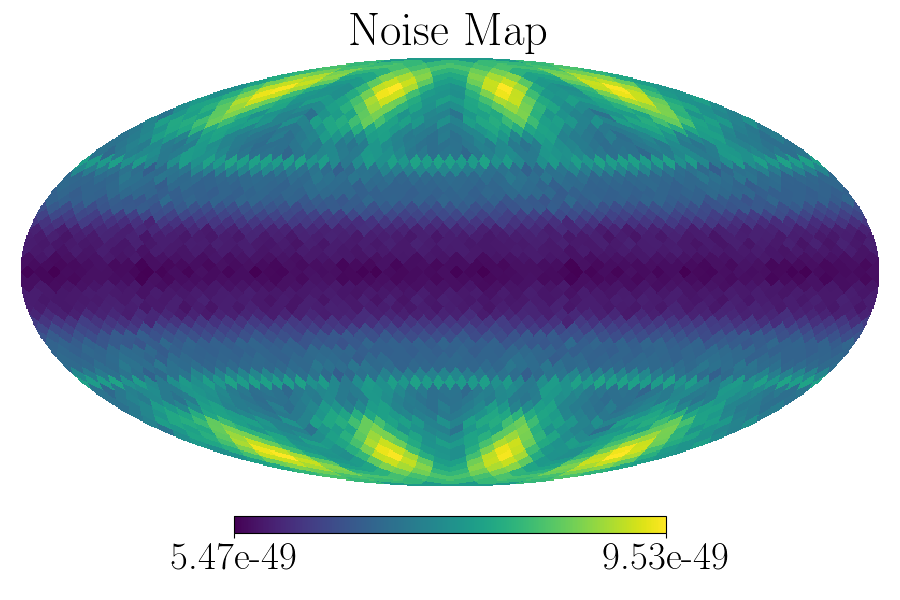}
    }
    \subfloat{
        \includegraphics[width=.29\textwidth]{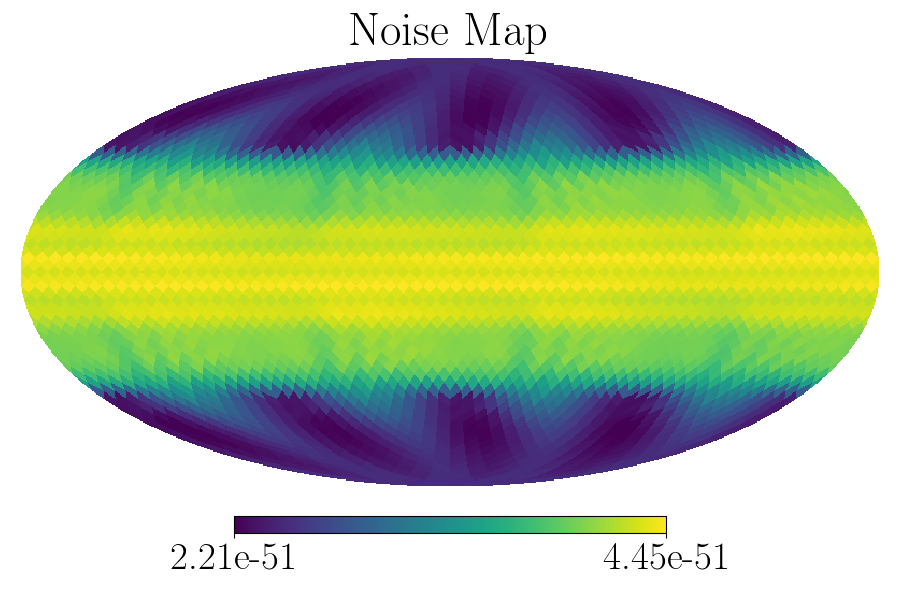}
    }

    \caption{Clean maps, SNR maps and noise maps broadband-integrated over $[20, 1726]$ Hz at a reference frequency of $f_\mathrm{ref}=25$ Hz using data from LIGO--Virgo's first three observing runs. From left to right are for spectral indices 0, $2/3$ and 3.}
    \label{fig:broadband_maps}
\end{figure*}
\begin{figure}[!ht]
    \centering
    \includegraphics[width=0.91\columnwidth]{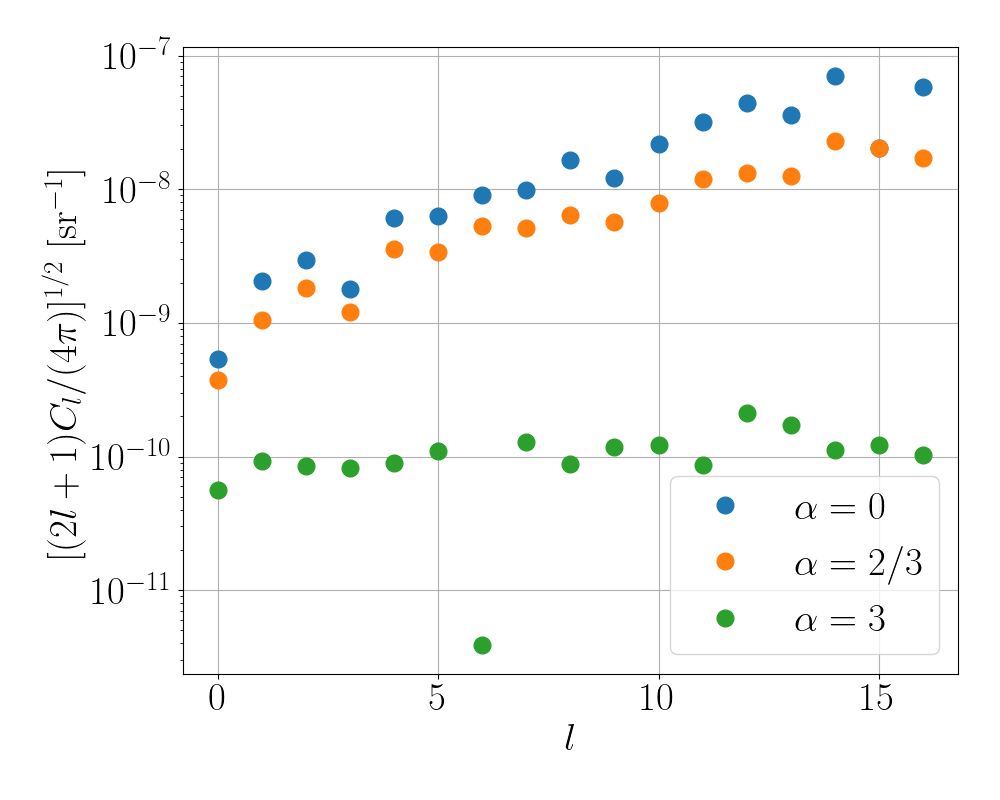}
    \caption{95\% upper limits on the angular power spectrum $C_l$ of the SGWB for power laws of $\alpha=0, \, 2/3$ and 3 at a reference frequency $f_\mathrm{ref}=25$ using data from LIGO--Virgo's first three observing runs. We have noted the outlier for $\ell=6$ in the $\alpha=3$ case: this is currently under investigation and is believed to be due to a noise fluctuation which makes the point value of $C_6$ negative.}
  \label{fig:O1-O3_Cls}
\end{figure}
In Fig.~\ref{fig:sim_mapping}, we show input maps, reconstructed clean maps, SNR maps and normalized residual maps for all three cases. All injected maps are successfully recovered, with minimal residual maps. We have also verified all combinations of injected maps and spectral indices not shown in Fig.~\ref{fig:sim_mapping}. Note that our mapping method in the pixel domain successfully recovers both extended sources as in the case of the monopole and Gaussian maps as well as the map with 30 random point sources, although with some caveats. The ``point'' sources are generated in the same resolution as the recovery map, meaning that each ``point'' here spans 13.4 deg$^2$. For a more realistic check, point source simulations need to be generated at a finer resolution and recovered by coarser graining. Also, for $\alpha=3$, the Fisher matrix is well-conditioned hence we do not need to apply conditioning in its inversion. Without the information loss, point sources for $\alpha=3$ are well recovered whereas for other spectral indices the recoveries manifest leakage and loss of resolution. An example of point source recovery for $\alpha=0$ is shown in Fig.~\ref{fig:randpoints_alpha0}, illustrating the ``smearing'' of the point source recovery. Limited by computational resources, we defer work on improving the resolution on the clean map to future work.

For spectral-model-independent narrowband searches, we show the pipeline's capability to probe spectral shapes using injected maps of a monopole, a Gaussian random field and 30 random point sources. The pipeline runs map-making in 30 adaptively chosen frequency bands in the search range of [20, 1726] Hz for each simulation. In each band, we set $f$ to be the midpoint frequency, and adaptively produce sky maps of 192, 768, 3072 pixels, or equivalently $N_\mathrm{side}=4,\,8,\,16$ in the \texttt{HEALPix} scheme, with each pixel covering 214.9 deg$^2$, 53.7 deg$^2$, 13.4 deg$^2$ respectively. %
These choices allow for good regularization of the Fisher matrix, and allow us to aptly over-resolve anisotropies according to the diffraction limit. %
We plot the reconstructed spectral shapes and energy densities in each case, along with the target model in Fig.~\ref{fig:sim_reconst_shape}. In the first two cases, map monopoles in different bands collectively trace out the expected spectral shape. The recovery of the spectrum is harder in the case of random point sources: we find that the monopole is not well recovered at lower frequencies, while the spectrum emerges in the higher frequency bands. As may be observed in Fig.~\ref{fig:sim_randpoint_maps}, the recovered maps at lower frequencies do not resolve the point sources, causing GW power leakage. More on this sort of effect is explained in~\cite{AR_Romano_phase}. This may also be due to a sub-optimal conditioning of the Fisher matrix, which can be explored by repeating simulations as described in Sec.~\ref{subsec:deconv} in each individual frequency band.

%% file: sections/5_LIGO-Virgo_data.tex
\section{Application to LIGO--Virgo Data} \label{sec:data}

We apply the methods outlined above to real data obtained by the LIGO and Virgo GW detectors. Our results clearly show no evidence for a signal, in agreement with the LVK results~\cite{O3_SGWB_anisotropic}, hence we set upper limits on anisotropies as well as the isotropic monopole as a limiting case using the maximum-likelihood mapping method in the pixel domain described in Sec.~\ref{sec:method}. We also set constraints on the spectral shape of the SGWB using the spectral-model-independent method described in Sec.~\ref{sec:gwb_model_indep}.

For the analyses, we use the publicly available folded datasets of the first three observing runs of Advanced LIGO and Advanced Virgo \cite{folded_data_O1-O3}. The strain time series is Fourier transformed and cross-correlated between each available pair of detectors in the network at the time of observing. The cross-correlated data from each pair are then folded over one sidereal day \cite{data_folding_fast, data_folding_veryfast}, reducing the computation time for anisotropic searches by a factor of the number of total observing days. This makes the processing of stochastic searches feasible in any modern-day personal computer. For O1 and O2, cross-correlated data only exist for the HL baseline, while for O3, data from all three combinations, HL, HV and LV, are available. Each sidereal-day folded dataset is chunked into 898 segments, with each segment lasting $\tau = 96$ s.

We perform all our analyses in the frequency range between 20 and 1726 Hz at a resolution of $1/32$ Hz, although 99\% of sensitivity for isotropic broadband analyses comes from the frequency band between 20 and 300 Hz \cite{O3_SGWB_isotropic}. This is because, depending on the spectral shape of the signal and the regularity of the Fisher matrix, anisotropic searches are not limited by the same sky-integrated sensitivity as isotropic searches.

\subsection{Spectral-model-dependent, Broadband Limits}

For the spectral-model-dependent, broadband searches, we present the results using three spectral indices, $\alpha=0, \, 2/3$ and 3, same as the LVK searches \cite{O3_SGWB_anisotropic}. The entire range of frequencies is integrated into a single map for each case of $\alpha$. Combining O1, O2 and O3 analyses, we show the reconstructed clean maps computed via Eq.~\eqref{eq:clean_map}, SNR maps via Eq.~\eqref{eq:snr_map} and noise maps via Eq.~\eqref{eq:noise_map} in Fig.~\ref{fig:broadband_maps} for each spectral index $\alpha=0, \, 2/3$ and 3 from left to right respectively. The condition number chosen for each index is listed in Table \ref{table:condition_num}. %

 We calculate the normalized GW energy density at a reference frequency of $f_\mathrm{ref}=25$ Hz for each spectral index and find these are consistent with 0, hence we set frequentist 95\% confidence level upper limits summarized in Table \ref{table:broadband_table}.
 Our upper limits are consistent with the LVK isotropic search results \cite{O3_SGWB_isotropic}.
From the SNR maps in Fig.~\ref{fig:broadband_maps} we find the maximum SNR values across the sky, reported in Table~\ref{table:broadband_table}. These are well below a significant deviation from 0.
To confirm this, we calculate p-values from the distributions of the SNR maps; these are also reported in Table~\ref{table:broadband_table}.
We thus conclude that we find no evidence of GW signals in either the monopole or anisotropies.
Note that the SNR maps are Gaussian distributed with norms less than 1: the same behaviour is observed in the LVK collaboration work~\cite{O3_SGWB_anisotropic}, and stems from the fact that the maps have been regularized. The p-values calculated here include this re-normalization.

We also show the upper limits on the angular power spectrum $C_l$'s of the SGWB obtained via Eq.~\eqref{eq:sgwb_cls} in Fig.~\ref{fig:O1-O3_Cls}. These are approximately consistent with the LVK anisotropic search results~\cite{O3_SGWB_anisotropic}, given that regularization is performed very differently, hence the spread over $\ell$ modes appears different in the two upper limits. 
Our choice of the maximum $\ell$ mode included here is dictated by our pixel resolution, jointly with the expected angular resolution of this style of search discussed in ~\cite{O3_SGWB_anisotropic}. %
The relation between $\ell$ mode and number of pixels necessary to resolve it, expressed in terms of the \texttt{HEALPix} $N_\mathrm{side}$ parameter, is roughly $\ell_{\rm max} \sim 2 N_\mathrm{side}$.
This would suggest going up to an $\ell_{\rm max} = 32$ for our analysis.
However, even in the most sensitive scenario ($\alpha=3$), according to~\cite{O3_SGWB_anisotropic} we expect resolutions higher than $\ell_{\rm max} > 16$ to be unattainable, due to the shape of the LIGO and Virgo noise curves.
Hence, we select $\ell_{\rm max} = 16$.

\begin{table}
\begin{center}
\begin{tabular}{>{\centering\arraybackslash}p{1.5cm} | >{\centering\arraybackslash}p{3cm} > {\centering\arraybackslash}p{3cm}}
  \multicolumn{1}{c|}{$\alpha$} & 
  \multicolumn{1}{c}{Max SNR (\% p-value)} & 
  \multicolumn{1}{c}{95\% upper limit on $\Omega_\alpha$} \\
 \hline
 0 & 1.9 (6) & $7.3 \times 10^{-9}$ \\ 
 $2/3$ & 2.2 (3) & $5.1 \times 10^{-9}$ \\ 
 $3$ & 2.7 (1) & $5.1 \times 10^{-10}$ \\
 \hline
\end{tabular}
\end{center}
\caption{\label{table:broadband_table}Maximum pixel SNRs of the reconstructed broadband clean maps and 95\% confidence level upper limits on the normalized GW energy density at a reference frequency of $f_\mathrm{ref}=25$ Hz for the HLV network using data from the first three observing runs.}
\end{table}

\subsection{Spectral-model-independent, Narrowband Limits}

\begin{figure*}
    \centering
    \subfloat{\label{a}\includegraphics[width=0.49\linewidth]{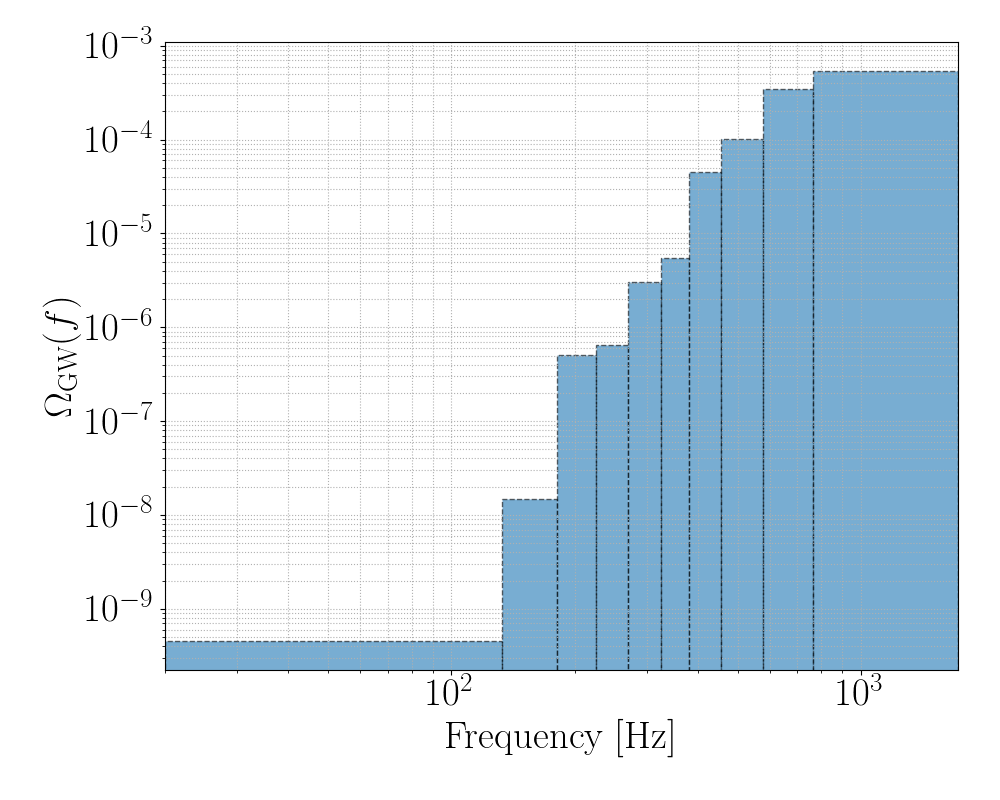}}
    \subfloat{\label{c}\includegraphics[width=0.49\linewidth]{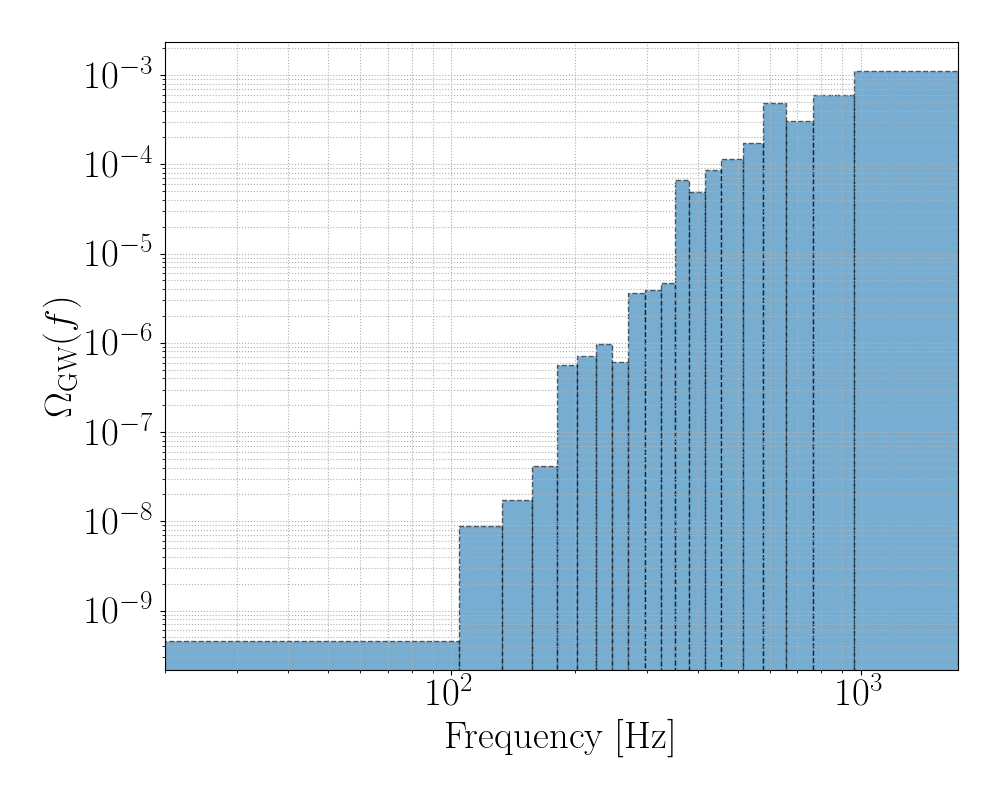}}
    
    \caption{95\% upper limits on the energy densities in distinct frequency bands used in the spectral-model-independent spectral analysis. We show our results in 10 and 20 bands. The method assumes a scale invariant spectral shape in each spectral band. The results are consistent with noise dominated estimates.}
    \label{fig:narrowband_omega}
\end{figure*}

Using the spectral-model-independent method described in Sec.~\ref{sec:gwb_model_indep}, we first divide the search range between 20 and 1726 Hz into 10 and 20 frequency bands with adaptively chosen endpoints. Since O3 achieves the best sensitivity out of the three observing runs and HL is the most sensitive out of the three baselines, the frequency endpoints are chosen such that each band contains the same amount of noise weighted strain power of the O3 HL data. O1 HL, O2 HL, O3 HV and O3 LV analyses then employ the same frequency banding as O3 HL. %

\begin{figure*}
    \centering
    \subfloat{
        \includegraphics[width=.3\textwidth]{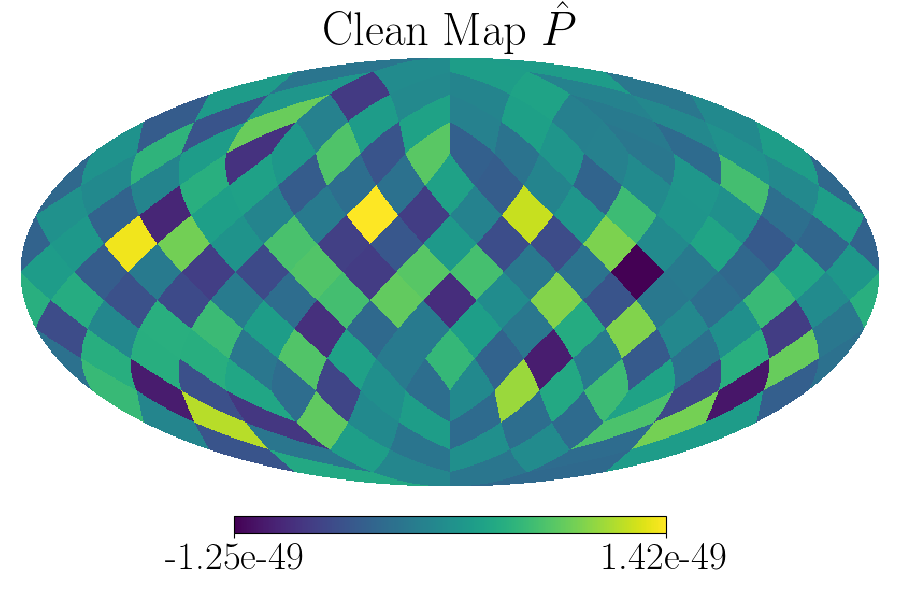}
    }
    \subfloat{
        \includegraphics[width=.3\textwidth]{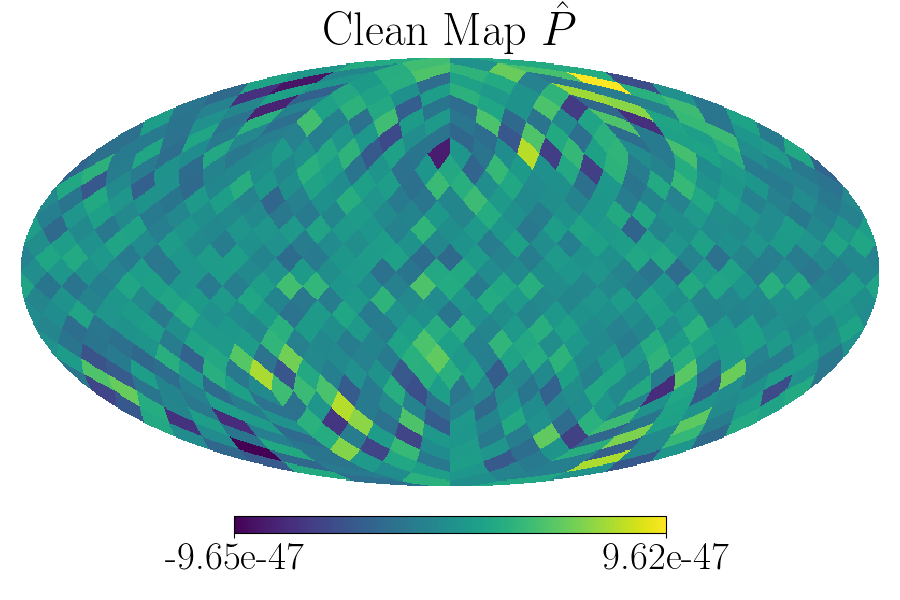}
    }
    \subfloat{
        \includegraphics[width=.3\textwidth]{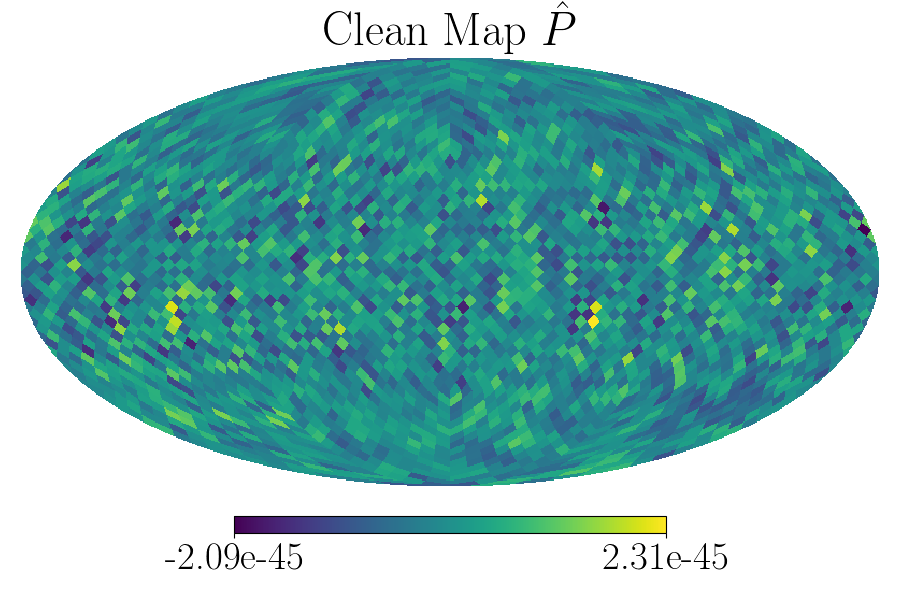}
    }

    \subfloat{
        \includegraphics[width=.3\textwidth]{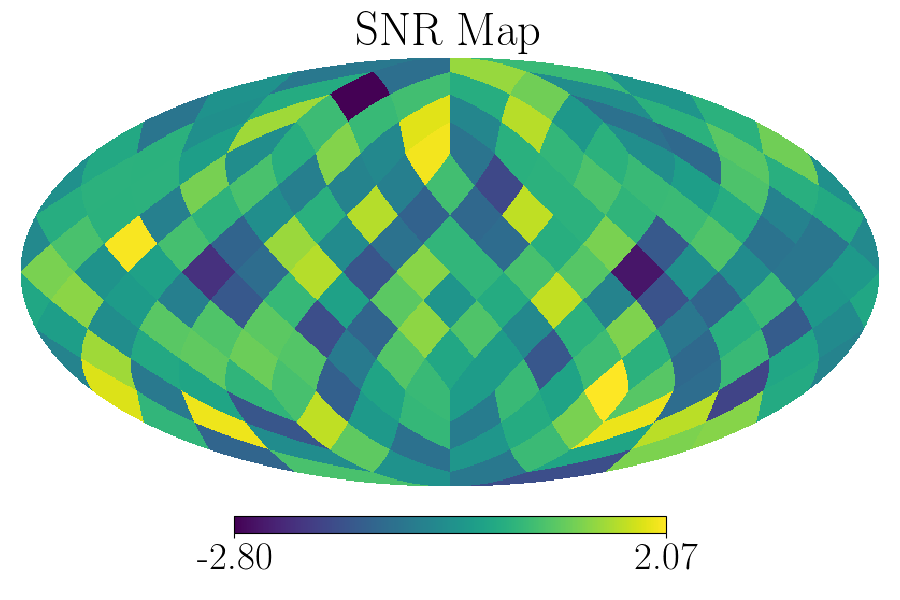}
    }
    \subfloat{
        \includegraphics[width=.3\textwidth]{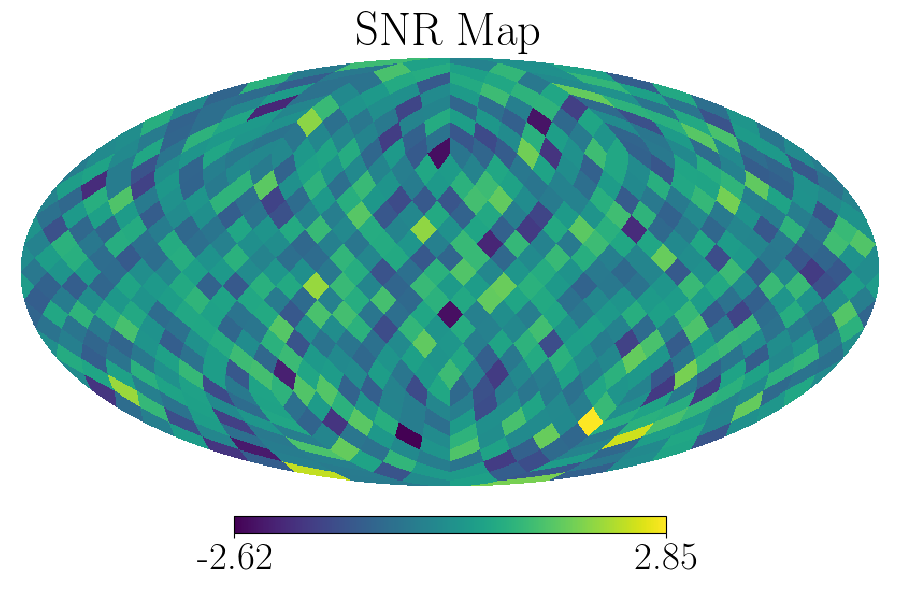}
    }
    \subfloat{
        \includegraphics[width=.3\textwidth]{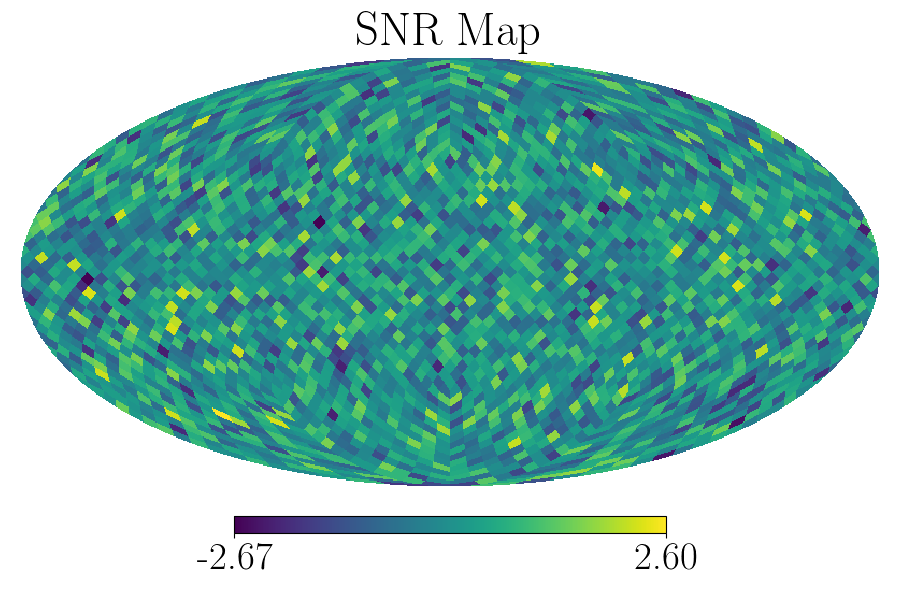}
    }

    \subfloat{
        \includegraphics[width=.3\textwidth]{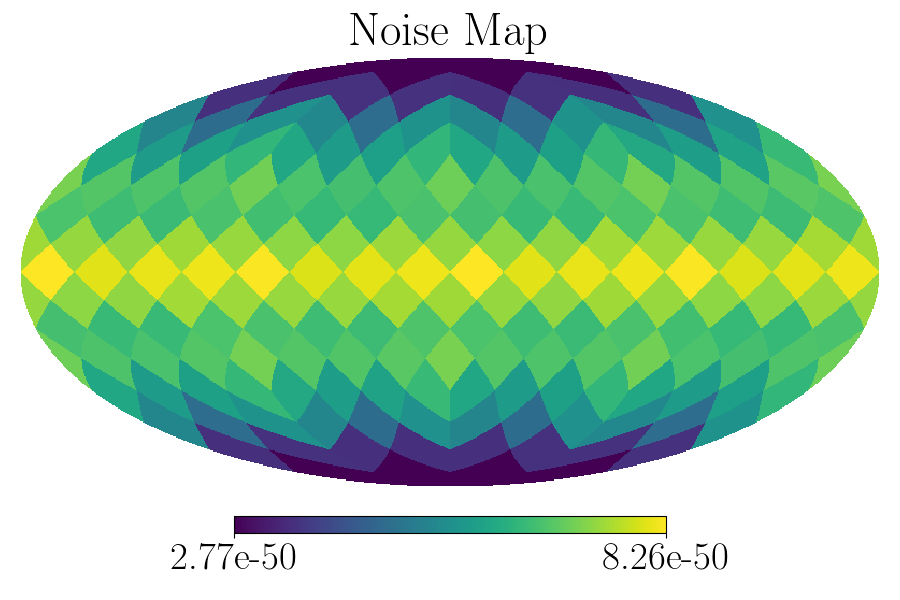}
    }
    \subfloat{
        \includegraphics[width=.3\textwidth]{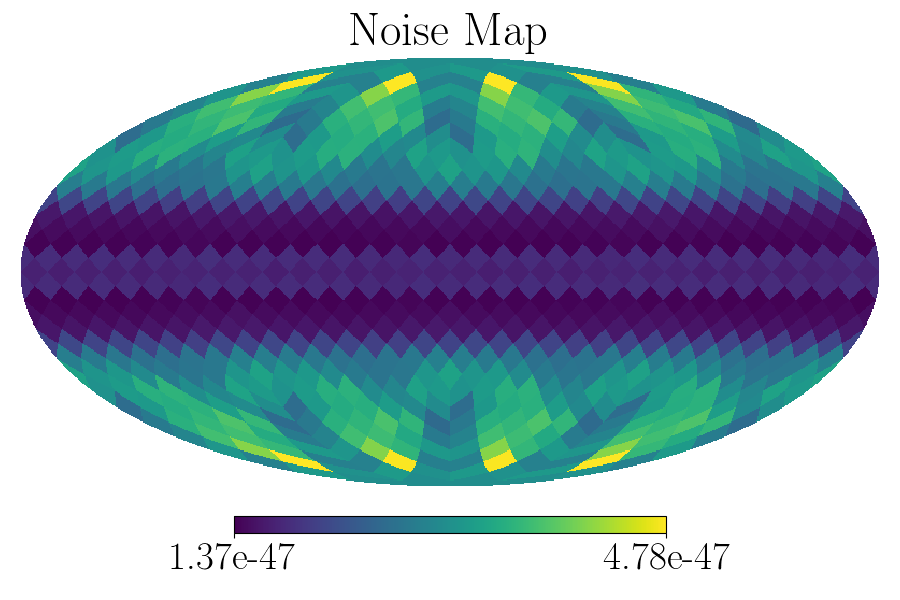}
    }
    \subfloat{
        \includegraphics[width=.3\textwidth]{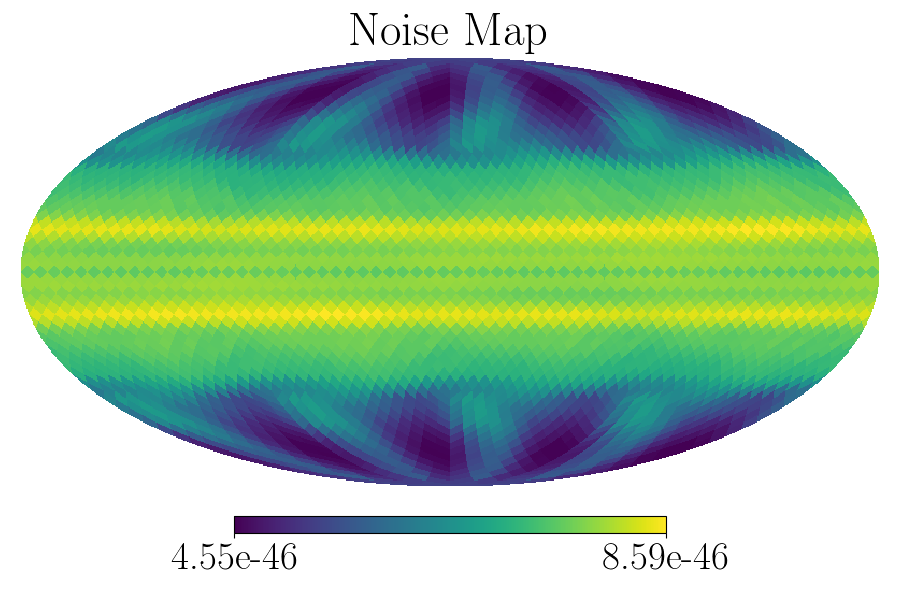}
    }

    \caption{Clean maps, SNR maps and noise maps for three frequency bands representative of low, mid and high frequencies in the spectral-model-independent search. In this analysis of the combined O1+O2+O3 data, the search range is divided into 10 bands. From left to right, the plots shown are for frequencies between 20 and 133.125 Hz, between 270.5 and 324.21875 Hz and between 765.8125 and 1726 Hz, and $N_{\rm pix} = 192$, $768$, $3072$ respectively.}
    \label{fig:narrowband_maps}
\end{figure*}

The mapping method described in Sec.~\ref{sec:gwb_max_likelihood} is run on each band separately and the resulting upper limits on the GW energy density are plotted in Fig.~\ref{fig:narrowband_omega}. The energy densities in different bands collectively probe the spectral shape of the SGWB. The spectral shapes obtained in our analyses are consistent with a noise-dominated estimate with increasing power as a function of frequency, resembling the detector noise curve. We also show the clean maps, SNR maps and noise maps for three narrowband analyses in low, mid and high frequencies of the 10-band case in Fig.~\ref{fig:narrowband_maps}. The lowest frequency band is between 20 and 133.125 Hz; the mid band is between 270.5 and 324.21875 Hz; and the highest band is between 765.8125 and 1726 Hz. We note that the changes in the scale of structures are evident as frequencies increase and our method chooses the resolution of each band accordingly as demonstrated in Fig.~\ref{fig:narrowband_maps}. 

%% file: sections/6_conclusions.tex
\section{Conclusions} \label{sec:conclusions}

In this work, we have developed a maximum-likelihood mapping method in the pixel domain for the SGWB power on the sky, complimentary to the methods of the LVK collaboration \cite{O3_SGWB_anisotropic}. In SGWB mapping, Fisher matrix regularization has long been an active area of research. We have presented an empirical method, albeit preliminary, to systematically regularize the Fisher matrix in mapping deconvolution via monopole simulations. In addition to modeled searches, we have introduced an improved spectral-model-independent, narrowband search method to probe the spectral shape of the SGWB, with adaptive frequency banding and adaptive pixelization techniques applied to each band. We have shown that this is a valid method to probe spectral shapes of anisotropic backgrounds, and may serve as a first step to characterize these signals which may then inspire a targeted search with a more refined model.

We have verified both the modeled and the unmodeled methods in various simulations and we apply both to LIGO--Virgo's folded datasets from the first three observing runs. In the spectral-model-dependent, broadband searches, we do not find any excess signals on top of the detector noise. In the spectral-model-independent, narrowband searches, our obtained spectral shapes are consistent with noise dominated estimates. Our results are in agreement with what is found by the LVK, as summarized in Table~\ref{table:broadband_table}.

In future work, we will improve the reliability of the Fisher matrix regularization method when applied to narrowband searches. The method is sub-optimal in narrow bands as the Fisher matrix conditioning will be band-dependent. This is particularly evident in our simulations of point sources, which are very sensitive to Fisher regularisation (see Figs.~\ref{fig:sim_reconst_shape} and~\ref{fig:sim_randpoint_maps}). Ideally, the condition number for each band is independently determined, while in this study we have used the broadband condition numbers as an alternative. 
Furthermore, we can explore setting constraints on different parametric models of the SGWB spectral shape starting from our spectral-model-independent results.
Finally, the ultimate goal of the spectral-model-independent method is to extend its capability to search for angular-dependent, frequency-dependent (most general) backgrounds. %

In expectation of a first detection of SGWBs in the coming observing runs, we also plan to use the pipeline to probe interesting questions. For example, we aim to assess whether we should expect to detect the isotropic or anisotropic component of the SGWB first, assuming different observing scenarios and signal characteristics. 

%% file: sections/7_acknowledgements.tex
\section*{Acknowledgements} \label{sec:acknowledgements}
This material is based upon work supported by NSF's LIGO Laboratory which is a major facility fully funded by the National Science Foundation. LIGO was constructed by the California Institute of Technology and Massachusetts Institute of Technology with funding from the National Science Foundation (NSF) and operates under cooperative agreement PHY-0757058. The authors are grateful for computational resources provided by the LIGO Laboratory and supported by NSF Grants PHY-0757058 and PHY-0823459. We would like to thank Jishnu Suresh, Deepali Agarwal, and Leo Tsukada for providing helpful comments and suggestions. L.X., A.I.R., and A.J.W. were supported by the NSF award 1912594. This paper carries LIGO Document Number LIGO-P2200339.